\documentclass[%
 aip,
 jcp, 
 nofootinbib,
 amsmath,amssymb,
 reprint,
]{revtex4-1}

\usepackage{graphicx}
\usepackage{dcolumn}
\usepackage{xcolor}
\usepackage{adjustbox}
\usepackage{import}
\usepackage{booktabs}
\usepackage{mathtools}
\usepackage{comment}
\usepackage{hyperref}
\usepackage{multirow}
\usepackage{array}
\usepackage{makecell}
\usepackage{siunitx}   
\usepackage[utf8]{inputenc}
\usepackage{textgreek}

\DeclarePairedDelimiter\bra{\langle}{\rvert}
\DeclarePairedDelimiter\ket{\lvert}{\rangle}
\DeclarePairedDelimiterX\braket[2]{\langle}{\rangle}{#1\,\delimsize\vert\,\mathopen{}#2}

\usepackage{algorithm}
\usepackage{algpseudocode}

\usepackage{bm}

\usepackage[utf8]{inputenc}
\usepackage[T1]{fontenc}
\usepackage{mathptmx}
\usepackage{etoolbox}

\newcommand{\citeme}[1][]{\textcolor{red}{cite}}

\DeclareUnicodeCharacter{2500}{\textemdash} 
\DeclareUnicodeCharacter{2212}{-}

\makeatletter
\def\@email#1#2{%
 \endgroup
 \patchcmd{\titleblock@produce}
  {\frontmatter@RRAPformat}
  {\frontmatter@RRAPformat{\produce@RRAP{*#1\href{mailto:#2}{#2}}}\frontmatter@RRAPformat}
  {}{}
}%
\makeatother
\begin{document}

\preprint{AIP/123-QED}

\title
{
Approximating Hartree-Fock theory via an efficiently local
reformulation
}

\author{Trine Kay Quady${}^\dagger$}
 \affiliation{ 
Department of Chemistry, University of California, Berkeley, California 94720, USA
}
\affiliation{%
Chemical Sciences Division, Lawrence Berkeley National Laboratory, Berkeley, California 94720, USA
}%

\author{Eric Neuscamman}
 \email{eneuscamman@berkeley.edu}

  \affiliation{ 
Department of Chemistry, University of California, Berkeley, California 94720, USA
}
\affiliation{%
Chemical Sciences Division, Lawrence Berkeley National Laboratory, Berkeley, California 94720, USA
}%

\date{\today}

\begin{abstract}

We explore a reorganized framework for the Hartree Fock equations
that allows varying patterns of locality to be imposed on the molecular 
orbitals while maintaining a highly efficient self-consistent field
optimization algorithm.
Rather than limiting orbitals' spread and then variationally minimizing
the energy within those limits, our reorganization neatly pairs each
local degree of freedom with a specific solution condition that itself has
a naturally local interpretation.
These pairs can each be turned on or off, and, regardless of the sparsity
pattern used to make such choices, the overall method maintains
a fast self-consistent field optimization.
We use this structure to test reaction-matched schemes for imposing
orbital locality and, through Fock builds that exploit this locality, 
achieve competitive timings even in modestly sized molecules.
Our initial tests also suggest that this approach to imposed
orbital locality can be arranged so as to do minimal damage
to both Hartree Fock and MP2 reaction energy predictions.

\end{abstract}

\maketitle

\section{\label{sec:level1}Introduction } 

Hartree Fock (HF) theory\cite{hartreeWaveMechanicsAtom1928,
fockNaeherungsmethodeZurLoesung1930} is an indispensable part of quantum chemistry's
methodological stack, but its computational cost is increasingly limiting
what quantum chemistry is able to do.
This problem exists both for very large molecules, where progress on
local correlation methods has increasingly made HF the computational
bottleneck,\cite{pinskiSparseMapsSystematic2015,
kopplParallel2016,
mesterBasisSetLimit2024,
szaboLinearScaling2023,
nagy_state---art_2024,
nagy_approaching_2019,
nagyOptimization2018,
saitowNewNearlinearScaling2017} 
and in more modestly sized molecules, where
training system-specific machine-learned interatomic potentials
can involve very large numbers of HF
(or, in practice, hybrid density functional)
calculations. \cite{unke2021machine}
One long-recognized route to reducing costs is to leverage
orbital locality, either whatever natural locality is
hiding just a unitary transform away from the canonical occupied
orbitals or an imposed locality gained at the price of approximating
the theory.
The latter route offers more in terms of cost savings but begins
to intrude on the standard self-consistent field (SCF) optimization
algorithm, as such local orbitals are usually not the eigenvectors
of any one Fock operator.
One consequence of this reality is that imposed locality schemes
typically have a hard time competing on cost with standard HF until
the system under study gets relatively large.
Here, we explore an alternative organization of
the HF equations that facilitates competitive optimization
efficiency even in modestly sized systems under significant
degrees of imposed locality.

Locality is often, though not always, invoked in approaches
seeking to lower the cost of HF theory's Fock build,
specifically the cost of constructing its
coulomb and exchange matrices.
Depending on the strategy used, simply localizing the
canonical occupied orbitals via a unitary  transformation\cite{pipekFastIntrinsicLocalization1989, lehtolaPipekMezeyOrbital2014, fosterCanonicalConfigurationalInteraction1960, edmistonLocalizedAtomicMolecular1963, vonniessenDensityLocalizationAtomic1972}
may or may not speed up the Fock build.
As many algorithms are based on HF's one-body density matrix,
\cite{helmich2021improved,merlot2013attractive,schwegler_linear_1997,li_density-matrix_1993}
which is invariant under unitary remixings of the occupieds,
they will not benefit from such localizations.
Algorithms that work directly in terms of the linear combination
of atomic orbital (LCAO) coefficients,
such as the local K \cite{aquilante2007low}
and occ-RI-K \cite{manzer2015fast}
exchange build methods, on the other hand,
can benefit substantially if localization allows them to
screen out large ranges of their index summations.
Similarly, localized orbitals are essential in
most local correlation methods.\cite{pulayOrbitalinvariantFormulationSecondorder1986,
saeboLocalTreatmentElectron1993,
neeseEfficientAccurateLocal2009, 
wernerScalableElectronCorrelation2015,
riplingerEfficientLinearScaling2013,
pinskiSparseMapsSystematic2015,
wangLocalSecondOrderMoller2023,
wangMoreNumericalPrecision2025,
wangSparsityElectronRepulsion2023,
maslenLocalitySparsityInitio1998,
maslenNoniterativeLocalSecond1998,
bangerterLowScalingTensorHypercontraction2021,
wangAutomaticConstructionInitial2019,
yangOrbitalspecificvirtualLocalCoupled2012,
kurashigeOptimizationOrbitalspecificVirtuals2012,
subotnikLimitsLocalCorrelation2008,
ayalaLinearScalingSecondorder1999,
ayalaAtomicOrbitalLaplacetransformed2001,
liEfficientImplementationClusterinmolecule2006,
maurerEfficientDistanceincludingIntegral2013,
schutzAnalyticalEnergyGradients2004,
schutzLoworderScalingLocal1999,
yangTensorFactorizationsLocal2011}. 
Whether a method benefits from sparsity in the
density matrix or from sparsity in the LCAO coefficients,
imposing additional locality on the orbitals beyond what can be
achieved through unitary remixings
offers the potential for even larger speedups,
albeit at the cost of introducing an additional layer
of approximation.

In practice, however, imposed locality methods tend to require
relatively large system sizes before they are advantageous.
In approaches that rely on fragmentation algorithms during
their main SCF cycle,
\cite{pengRegularizedLocalizedMolecular2022,nakai2023divide,
kitauraFragmentMolecularOrbital1999,zhangMolecularFractions2003}
the overhead involved typically does not pay off until
one reaches large systems.
Approaches that variationally minimize the energy
under an imposed locality scheme typically trade HF's
highly-efficient, DIIS-accelerated, Roothaan-style
SCF solver for a Newton or quasi-Newton direct minimization.
\cite{mrovecHowDelocalizedAre2024,burton_geometric_2025,aslattery_economical_2024,stoll_use_1980}
In practice, such methods typically require
a larger number of Fock builds to converge.\cite{vanVoorhisAGeometric2002,aslattery_economical_2024}
Ideally, one would prefer to achieve the benefits of
imposed locality while retaining the efficiency of
the traditional low-overhead SCF algorithm.

In the present study, we investigate an approach to HF
theory that associates each orbital degree of freedom with
a specific, chemically interpretable solution condition
in a way that allows locality to be imposed while
maintaining a low-overhead, DIIS-accelerated SCF cycle.
The basic idea will be threefold: (i) to formulate a way to
impose the zeroing out of the occupied-virtual block
of the Fock matrix without actually solving for the
virtual orbitals,  (ii) to render this formulation in
a local basis that allows each individual solution condition
to pair naturally with one orbital degree of freedom, and
(iii) to combine these conditions with similarly facile
conditions for orthonormality and the maximization
of an orbital localization function.
Crucially, even when large fractions of the orbital degrees
of freedom and their corresponding solution conditions
are switched off to achieve imposed locality,
the remaining equations can still be solved by a
DIIS-accelerated SCF cycle within which the traditional
Fock matrix diagonalization is replaced by a similarly low-overhead
inner kernel.
The approach can be paired with any Fock build strategy that
benefits from unusually local orbitals, and although
resolution of the identity (RI) approaches will likely be
the most efficient way to exploit such locality, our
preliminary data shows that a Cholesky-based approach is
already highly competitive.

To test the usefulness of this approach to imposed
locality, we combine it with a reaction matching strategy
that imposes increasingly severe locality constraints on orbitals
farther away from a reaction center.
In light of recent work by Mrovec and Gill \cite{mrovecHowDelocalizedAre2024}
that has shown how small the energetic consequences of imposed
locality can be even in highly delocalized $\pi$ networks,
we test our approach on isomerization reactions in highly
conjugated molecules.
We find that reaction energies within a kcal/mol of canonical
HF and MP2 theory can be achieved even when about half of
the LCAO coefficients are disabled in systems with as few
as 12 second row atoms.
Indeed, our preliminary timing data suggests that the approach
offers real computational advantages even at such
modest system sizes, confirming that low-overhead SCF
efficiency is being achieved despite the imposition of
a spatially heterogeneous set of locality constraints.

\section{Theory}

\subsection{A Local Subset of the Hartree-Fock Equations}
\label{sec:local_subset}

In HF theory, we seek to
make the energy of the Slater determinant stationary
under the constraint that the molecular orbitals are
orthonormal.
To achieve this goal, it is
sufficient to solve the following equations.
\begin{align}
    \label{eqn:BTFA}
    \mathbf{B}^\top \mathbf{F} \mathbf{A} &= 0 \\
    \mathbf{A}^\top \mathbf{S} \mathbf{A} &= \mathbf{I} \\
    \mathbf{B}^\top \mathbf{S} \mathbf{A} &= 0 \\
    \label{eqn:BTSB}
    \mathbf{B}^\top \mathbf{S} \mathbf{B} &= \mathbf{I}
\end{align}
Here, $\mathbf{A}$ and $\mathbf{B}$ contain the LCAO
coefficients for the occupied and virtual MOs, respectively,
$\mathbf{S}$ is the AO overlap matrix, and $\mathbf{F}$
is the Fock matrix in the AO basis.
To arrive at working equations for our local reformulation,
we will proceed through a series of manipulations that
bring us into a form in which individual elements of the
matrix equations can be associated with local degrees
of freedom, some of which should be safe to discard without
unduly impacting the accuracy of reaction energies
and other energy differences.

First, we will parameterize the virtual orbitals via a
linearly independent set of rough local virtuals (RLV).
The details of how we construct the RLVs are provided
below, but, for now, let us focus on their key features.
Each RLV will live on a small set of adjacent atoms, making
the LCAO matrix $\tilde{\mathbf{B}}$ that defines the RLVs
sparse:  each of its columns will have nonzero
elements in the rows of only $O(1)$ atoms.
We repeat this setup for the occupied space as well,
defining a set of rough local occupied (RLO) orbitals
which are linearly independent from each other
and from the RLVs.
Note, however, that none of these rough local orbitals
are required to be orthogonal to each other, and,
in practice, they will not be in order to allow them
to be more localized than orthogonality would permit.
Like the RLVs, each RLO lives on a small set of
adjacent atoms, ensuring that the columns of
the corresponding LCAO matrix $\tilde{\mathbf{A}}$
each contain only $O(1)$ nonzero elements.
Using these rough orbitals, we parameterize the
LCAO matrix for our final occupied orbitals
as follows.
\begin{align}
    \mathbf{A} =   \tilde{\mathbf{A}} (\mathbf{I} + \mathbf{U})
                 + \tilde{\mathbf{B}} \mathbf{V}
\end{align}
Our underlying variables are thus the elements
of the $n_o$ by $n_o$ matrix $\mathbf{U}$ and the
the $n_v$ by $n_o$ matrix $\mathbf{V}$,
where $n_o$ and $n_v$ are the numbers of occupied
and virtual orbitals, respectively.
So long as the columns of $\tilde{\mathbf{A}}$
and $\tilde{\mathbf{B}}$ form a set of
linearly independent orbitals,
this parameterization of
$\mathbf{A}$ is fully general and we retain the
same variational flexibility as if we had simply
chosen our parameters to be the elements of
$\mathbf{A}$ itself.
What we have gained is a setup where each
variable is tied to the chemically
intuitive contribution that an individual RLO or RLV
makes to one of our final occupied MOs.
Ultimately, we will use this intuition to disable most
of the elements of $\mathbf{U}$ and $\mathbf{V}$ to
produce an approximate theory.
Before making this approximation, however, it is
instructive to rewrite standard HF theory's working
equations in terms of this setup.

Defining the projector onto the orthogonal compliment of
the occupied span as
\begin{align}
    \mathbf{P} = \mathbf{I} - \mathbf{A} \mathbf{A}^\top \mathbf{S},
\end{align}
we can rewrite our HF equations as follows.
\begin{align}
    \label{eqn:modified_brillouin}
    \tilde{\mathbf{B}}^\top \mathbf{P}^\top \mathbf{F} \mathbf{A} &= 0 \\
    \mathbf{A}^\top \mathbf{S} \mathbf{A} &= \mathbf{I}
    \label{eqn:repeat_ATSA}
\end{align}
So long as our set of RLVs is linearly independent from the final
HF occupied orbitals, the solution to these equations produces
the same Slater determinant as we would have gotten had we
solved Eqs.\ (\ref{eqn:BTFA} - \ref{eqn:BTSB}).
In fact, both sets of equations have an infinite manifold of
solutions, as any unitary remixing of the final occupied
orbitals will also be a solution.
In canonical HF, we pick one specific solution within this
manifold by insisting that the orbitals be eigenvectors
of the Fock operator, but in the present approach this
condition is undesirable as it works against locality.
Instead, we will pick a specific solution by demanding
that we choose the unitary remixing of the occupied
orbitals that extremizes the objective function
$\mathcal{L}(\mathbf{U})$
of an orbital localization approach (e.g., the Pipek
Mezey objective function could be used).
Thus, instead of the Fock eigenvalue condition, we will
append our system of equations with the condition
\begin{align}
    \label{eqn:ddx}
    \mathbf{Q} = 0 \qquad \qquad
    \mathbf{Q} \equiv \left. \frac{\partial}{\partial \mathbf{X}}
    \Bigg(\mathcal{L}\Big(\mathbf{U} e^{\mathbf{X}} \Big) \Bigg)
    \right |_{\mathbf{X}=0}
\end{align}
where $\mathbf{X}$ is an anti-Hermitian orbital rotation matrix.
Noting that Eq.\ (\ref{eqn:modified_brillouin}) has $n_v n_o$
distinct elements, and that Eq.\ (\ref{eqn:repeat_ATSA})
and Eq.\ (\ref{eqn:ddx}) have $n_o (n_o + 1) / 2$
and $n_o (n_o - 1) / 2$ unique elements, respectively,
we see that the system of equations formed by
Eqs.\ (\ref{eqn:modified_brillouin} - \ref{eqn:ddx})
has one equation
for each element of $\mathbf{U}$ and $\mathbf{V}$.

Thus, if we wanted to, we could find the HF determinant
by designing an algorithm to solve these equations rather
than the usual Roothaan-style self-consistent field (SCF) equations.
Typically this is not the approach taken, because the
Roothaan approach is highly efficient:
aside from the Fock build itself, it requires one
$O(n^3)$-cost matrix diagonalization per SCF iteration.
We will want to ensure that solving
Eqs.\ (\ref{eqn:modified_brillouin} - \ref{eqn:ddx})
amounts something similarly efficient, but, first,
we should emphasize that, so far, we have not
actually made any approximations.
Indeed, all we have done is rearrange the math a bit,
but we have done so to achieve a specific goal:
we now have a system of equations for HF theory
in which the difficult-to-localize canonical virtual
orbitals (or any exact virtuals for that matter) are not
present, and in which the underlying variables in
$\mathbf{U}$ and $\mathbf{V}$ have useful local interpretations
that allow them to be straightforwardly paired with specific
solution conditions.

To see how, imagine a molecule that has a C-H bond
on one end of it, and a C-F bond on the other end.
If these ends are not too close to each other, then we would
expect that the
coefficients within $\mathbf{U}$ and $\mathbf{V}$
that mix the RLOs and RLVs from the C-F moiety into the final MO
for our localized C-H $\sigma$ bond will not be very important.
Likewise, the element in $\mathbf{U}$ that mixes the C-H bond's
RLO into the final MO for the C-F bond will also probably not
be particularly important.
If we wanted to approximate the theory by neglecting these
coefficients, the setup we've arrived at offers a straightforward
choice for which equations could be set aside to keep the
number of variables and equations balanced.
Specifically, we could (a) choose not to enforce orthogonality
between the final MOs for these two $\sigma$ bonds, (b)
set aside the equation for $\mathcal{L}$ being extremized by
their mixing, and (c) neglect the Brillioun-like
conditions between the RLVs on one moiety and the
final occupied MOs on the other.
The aforementioned elements of $\mathbf{U}$ somewhat
naturally correspond to these orthogonality and localization
equations, and those of $\mathbf{V}$ to the elements of the
Brillouin conditions.
Thus, whatever mechanism we settle on for choosing a
sparsity pattern within $\mathbf{U}$ and $\mathbf{V}$,
we have a straightforward way of cutting away
equations to arrive at a final sparse
set of working equations that is equal in number to
our remaining variables and can thus be solved to determine
their values.

The core hypothesis of our approach is that, when
evaluating energy differences, the approximation introduced
by removing from standard HF the equations and variables
that correspond to mixing far-away RLOs and RLVs into a
particular MO should be small.
In exchange for tolerating this approximation, we should
be rewarded with MOs that are significantly more local
than would be possible were we enforcing all of the
orthogonality constraints.
Indeed, it is those constraints in particular that tend
to force MOs to have long-range tails.\cite{hoyvikCharacterizationGenerationLocal2016,
pengRegularizedLocalizedMolecular2022,
feng_efficient_2004,
peng_effective_2013}
With those conditions and their corresponding
variables set aside, we should arrive at more local
orbitals and all of the algorithmic benefits that come
with them.
Although it is only one example of such a benefit,
the Fock build itself should become significantly
less expensive, allowing us to attack the core
driver of HF theory's computational cost.
As we will show in our results below, this effect is
significant enough to be helpful even in
molecules of modest size.
To get there, however, we must finish specifying a
concrete theory.
This section has provided the basic framework, but, to
realize it in practice, many decisions must be made.
How might we choose the RLOs and RLVs?
How should we set the sparsity pattern in
$\mathbf{U}$ and $\mathbf{V}$?
What type of Fock build approach would benefit
especially from unusually local orbitals?
Finally, since we are explicitly avoiding the
standard SCF diagonalization step, how do we go
about solving these equations, anyways?
We will tackle these questions one at a time,
starting with the last one.

\subsection{Self-Consistent Field Equations}
\label{sec:scf}

As in the traditional SCF approach to HF theory, we will adopt
an outer SCF macro-cycle that updates the Fock matrix
and an inner solver that, holding the Fock matrix fixed,
solves for the orbital variables, which for us are the
elements of $\mathbf{U}$ and $\mathbf{V}$.
As discussed above, the matrix equations
Eqs.\ (\ref{eqn:modified_brillouin} - \ref{eqn:ddx})
contain one individual equation for each 
element of $\mathbf{U}$ and $\mathbf{V}$.
To formulate our inner solver, it is helpful to first
make the matching between variables and equations more explicit.
\begin{align}
\label{eqn:assoc_Vai}
V_{ai} \hspace{10mm} &\longleftrightarrow \hspace{6mm}
[\tilde{\mathbf{B}}^\top \mathbf{P}^\top \mathbf{F} \mathbf{A}]_{ai} = 0 \\
\label{eqn:assoc_Uii}
U_{ii} \hspace{10mm} &\longleftrightarrow \hspace{6mm}
[\mathbf{A}^\top \mathbf{S} \mathbf{A}]_{ii} - 1 = 0 \\
\label{eqn:assoc_Uij}
U_{ij}~,~U_{ji} \hspace{5mm} &\longleftrightarrow \hspace{6mm}
[\mathbf{A}^\top \mathbf{S} \mathbf{A}]_{ij} = 0 ~ , ~ Q_{ij} = 0
\end{align}
Here, $V_{ai}$ is associated with the corresponding element
of our modified Brillouin condition, the diagonal element
$U_{ii}$ is associated with with the corresponding
normalization condition,
and finally the $i\ne j$ pair of variables
$U_{ij}$ and $U_{ji}$ is associated with a pair made up of
one orthogonality equation and one localization equation.
With the Fock matrix held fixed, each of these equations
is a polynomial equation in our variables,
with polynomial orders of 3 for Eq.\ (\ref{eqn:assoc_Vai}),
2 for Eq.\ (\ref{eqn:assoc_Uii}),
and 2 and 4 for the pair in Eq.\ (\ref{eqn:assoc_Uij})
(assuming that a quartic Pipek-Mezey-like objective function
is used for localization).
In practice, we will be keeping $\mathbf{U}$ and $\mathbf{V}$ sparse,
and so most of the variables and their associated equations will be
set aside and neglected, but the ones that remain form
a system of polynomial equations
\begin{align}
    \mathcal{P}_k=0 \qquad \qquad k \in[1,2,\ldots,n_r]
\end{align}
where each $\mathcal{P}_k$ is one polynomial in the retained variables
and $n_r$ is the number of such variables.

To solve these equations, we will exploit the idea that our rough local
orbital guesses are pretty good, which would mean that the
elements of $\mathbf{U}$ and $\mathbf{V}$ should be small in magnitude.
Thus, if we adopt a notation in which $\mathcal{P}^{(m)}_k$ is $m$th
order piece of the $k$th polynomial, we can write each equation as follows.
\begin{align}
    \label{eqn:funny_Ax_equals_b}
    \mathcal{P}^{(1)}_k =
  - \mathcal{P}^{(0)}_k
  - \mathcal{P}^{(2)}_k
  - \mathcal{P}^{(3)}_k
  - \mathcal{P}^{(4)}_k
\end{align}
If the variables themselves are indeed small in magnitude, then the
$\mathcal{P}^{(0)}_k$ and $\mathcal{P}^{(1)}_k$ terms will
be the most significant, and we will find ourselves looking at
a system of \textit{almost} linear equations.
To solve them, we therefore adopt an \textit{almost} linear
Krylov subspace solver in the form of a modified generalized
minimal residual (GMRES) algorithm\cite{gmres} with block-diagonal preconditioning.
The right hand side of Eq.\ (\ref{eqn:funny_Ax_equals_b})
plays the role of the $\vec{b}$ vector in a standard
$\mathbf{M} \hspace{0.5mm} \vec{x} = \vec{b}$ setup,
while the coefficients within the $\mathcal{P}^{(1)}_k$ terms
provide the elements of $\mathbf{M}$.
As in standard GMRES, we compute the right hand side
at the beginning of the algorithm,
but, unlike in the standard method, we update it after every
handful of Krylov subspace iterations using GMRES's
new estimate of the variables.
Each time we perform this update, we must take new inner products
between the history of Krylov vectors and the $\vec{b}$ vector,
but doing so is cheap thanks to the sparse nature of our variable set.
For preconditioning, we set up an approximation to the diagonal of the
linear transformation matrix $\mathbf{M}$ that assumes the rough local
orbitals are actually orthonormal (they are not, but this simplification
proves accurate enough for preconditioning).
Under this assumption, the diagonal entries for the
linear terms in Eq.\ (\ref{eqn:assoc_Vai}) boil down to
differences between MO-basis Fock diagonal elements, while the diagonal
entries for Eq.\ (\ref{eqn:assoc_Uii}) are simply equal to two.
For each variable pair from Eq.\ (\ref{eqn:assoc_Uij}), the preconditioner
retains the corresponding $2\times2$ on-diagonal block of $\mathbf{M}$.
Outside these $1\times1$ and $2\times2$ blocks the preconditioner is
set to zero, making it trivial to invert.
In all other respects, the approach is standard preconditioned GMRES,
which we find typically converges tightly enough after a few
dozen iterations to provide useful updates for 
$\mathbf{U}$ and $\mathbf{V}$ for the next macro iteration of the
outer SCF cycle.
As in standard HF, this outer SCF cycle employs DIIS\cite{pulay_convergence_1980,
pulay_improved_1982}
to accelerate convergence (the retained elements of
Eq.\ (\ref{eqn:modified_brillouin}) are used as the DIIS error vector).

Thanks to the sparsity of the variable set, the cost of the matrix
vector multiply operations and the $\vec{b}$ vector constructions
within GMRES are cheap compared the Fock build.
If no sparsity were employed, they would amount to a bunch
of $O(n^3)$-cost dense matrix multiplies in close analogy to the
level-3 BLAS nature of standard HF's matrix diagonalization
(although likely with a larger prefactor).
With sparsity, they should scale as $O(n^2)$ or better,
and indeed the GMRES solve proves to be much less expensive than
the Fock build even in small molecules.
So, as in standard HF, the key cost bottleneck is the Fock build itself
and the related handling of the two-electron integrals.
Thanks to the extra-local nature of the occupied orbitals we produce,
we are particularly well set up for a Cholesky-decomposition-based
local orbital Fock build.

\subsection{Cholesky-Based Fock Build}
\label{sec:fock_build}

To construct the coulomb matrix $\mathbf{J}$
and exchange matrix $\mathbf{K}$ needed for $\mathbf{F}$,
we rely on a Cholesky factorization of the two-electron integrals.
\begin{align}
  (pq|rs) \approx \sum_\mu L^{\mu}_{pq} L^{\mu}_{rs}
\end{align}
For this study, we have employed the method of Koch and coworkers
\cite{folkestad2019efficient}
to prepare $L$.
Thanks to the fact that only $O(n)$ pairs $pq$ of AOs
yield integrals with appreciable magnitude, the range of the
Cholesky index $\mu$ can be expected
to be proportional to the number of AOs, and the total number
of nontrivial elements within $L$ will only grow as $O(n^2)$.
To achieve this memory scaling in practice, we implement $L$
in a block-sparse fashion, with the $\mu$, $p$, and $q$ index ranges
each grouped into chunks.
To maintain reasonable matrix matrix multiplication efficiency later
on, we break up the $\mu$ range into chunks of no smaller than 16,
with the final
chunk padded with zeros if the number of above-threshold Cholesky
vectors is not an even multiple of 16.
For the $p$ and $q$ ranges, each chunk
consists of all the AOs on a particular non-hydrogen
atom as well as those on any hydrogens bonded to that atom.
Thus, in for example the cc-pVDZ basis, $L$ is stored as a
sparse collection of dense tiles that are each roughly
$16 \times 25 \times 25$ elements in size
(the latter two dimensions
vary based on how many hydrogens are involved).
Note that breaking the $\mu$ range into chunks,
and not just the $p$ and $q$ ranges, is important,
because the magnitudes of the elements within $L$ get
smaller with increasing $\mu$, allowing larger and
larger fractions of the terms to be skipped in the
block sparse approach we now embark on as one
marches down the list of $\mu$ chunks.

To construct $\mathbf{J}$ and $\mathbf{K}$, we take an approach
similar to the local exchange method of Lindh and coworkers.
\cite{aquilante2007low}
To start, we form a set of half-transformed Cholesky factors
\begin{align}
    \label{eqn:R_from_L_A}
    R^\mu_{xj} = \sum_y L^\mu_{xy} A_{yj}
\end{align}
in which $\mathbf{A}$ is the occupied orbital LCAO matrix
from Section \ref{sec:local_subset}.
In practice, we do not evaluate $R$ exactly, but instead
aim to build a block-sparse approximation of it where the
neglected pieces are guaranteed to have contained no values
larger than the Cholesky threshold and where the elements
in the retained pieces are guaranteed to differ from their
true values by no more than the Cholesky threshold.
To achieve this outcome, we will break $R$ into pieces,
each of which covers one chunk of the $\mu$ range,
one chunk of the $x$ range, and the subset of the $j$
range that has above-threshold values for that particular
slice of the $\mu$ and $x$ ranges.
If we enumerate the $\mu,x$ pairs of one such piece of $R$
via the compound index $c$, then we can choose a
$j$ subset that safely captures all of that piece's
above threshold elements by, for each chunk of $y$,
finding the largest 1-norm among those partial rows of
the matrix $L_{cy}$ and combining them with the 1-norms of
the corresponding partial columns of $A$.
These norms allow an upper-bound to be placed on the elements
in this piece of $R$ for each $j$ value, and thus the $j$
values for which this upper-bound is below threshold to
be skipped.
In practice, storing tables of descending-magnitude 1-norms
for the partial $A$ columns for each $y$ chunk allows this
analysis to be carried out in $O(n^2)$ time (constant time
for each piece of $R$, with a quadratic number of pieces
due to the number of $\mu$-chunk$\hspace{0.3mm},x$-chunk pairs).
Once we have identified the $O(1)$-sized subset of the $j$
values that need to be retained for a given piece of $R$,
a similar 1-norm-based analysis allows us to bound the
individual $y$ chunks' contributions to that piece.
We set aside and neglect the largest set of such chunks
whose bounds add up to less than the Cholesky threshold,
which in practice shrinks the sum over $y$ for that
piece of $R$ to involve only $O(1)$ chunks.
Again, careful pre-tabulation allows this analysis to
be done at an $O(1)$ cost per piece of $R$.
All told, we produce a block sparse version of $R$
that takes up $O(n^2)$ memory in an amount of time that
also grows as $O(n^2)$.

With block-sparse versions of $L$ and $R$ in hand, evaluating
$\mathbf{J}$ and $\mathbf{K}$ is relatively straightforward.
\begin{align}
    J_{pq} &= \sum_{\mu x j} L^\mu_{pq} R^\mu_{xj} A_{xj} \\
    K_{pq} &= \sum_{\mu j} R^\mu_{pj} R^\mu_{qj}
\end{align}
In both cases, the cost of these sums grows quadratically
with system size.
In the case of $\mathbf{J}$, this follows from the fact
that, for each of the $O(n^2)$ pieces of $R$,
there are only $O(1)$ retained values for $j$, and,
after the $x$ and $j$ sums are complete, there are for
each chunk of $\mu$ only $O(n)$ nontrivial $pq$ pairs.
In the case of $\mathbf{K}$, the locality of the MOs
guarantees that, within each $\mu$ chunk,
a given value of $j$ will only have been retained by 
$O(1)$ of the $p$ range chunks and $q$ range chunks.
With $O(n)$ $\mu$ chunks and $O(n)$ $j$ values, 
the sparse summation to form $\mathbf{K}$ therefore also
has a quadratic cost.
Thanks to the block sparse approach, the actual
calculations take the form of a series of level 2 and
level 3 BLAS operations.

It is worth pointing out, as already highlighted by
Lindh and coworkers, \cite{aquilante2007low}
that the same basic strategy should also be achievable
in RI approaches that rely on pre-defined auxiliary
basis sets.
Indeed, if the history of Cholesky vs RI methods is
any indication, such an RI approach is likely to be
even more efficient.
In the present study, however, we satisfy ourselves
with the Cholesky approach, in part because we already
had most of what it needed in one of our code bases,
and in part because Cholesky provides a clean threshold with
which to control how much of the error we see is coming
from the integral decomposition vs the locality
approximation, which has been helpful during testing
and debugging.

\begin{figure}[]
    \centering
    \includegraphics[width=0.95\linewidth]{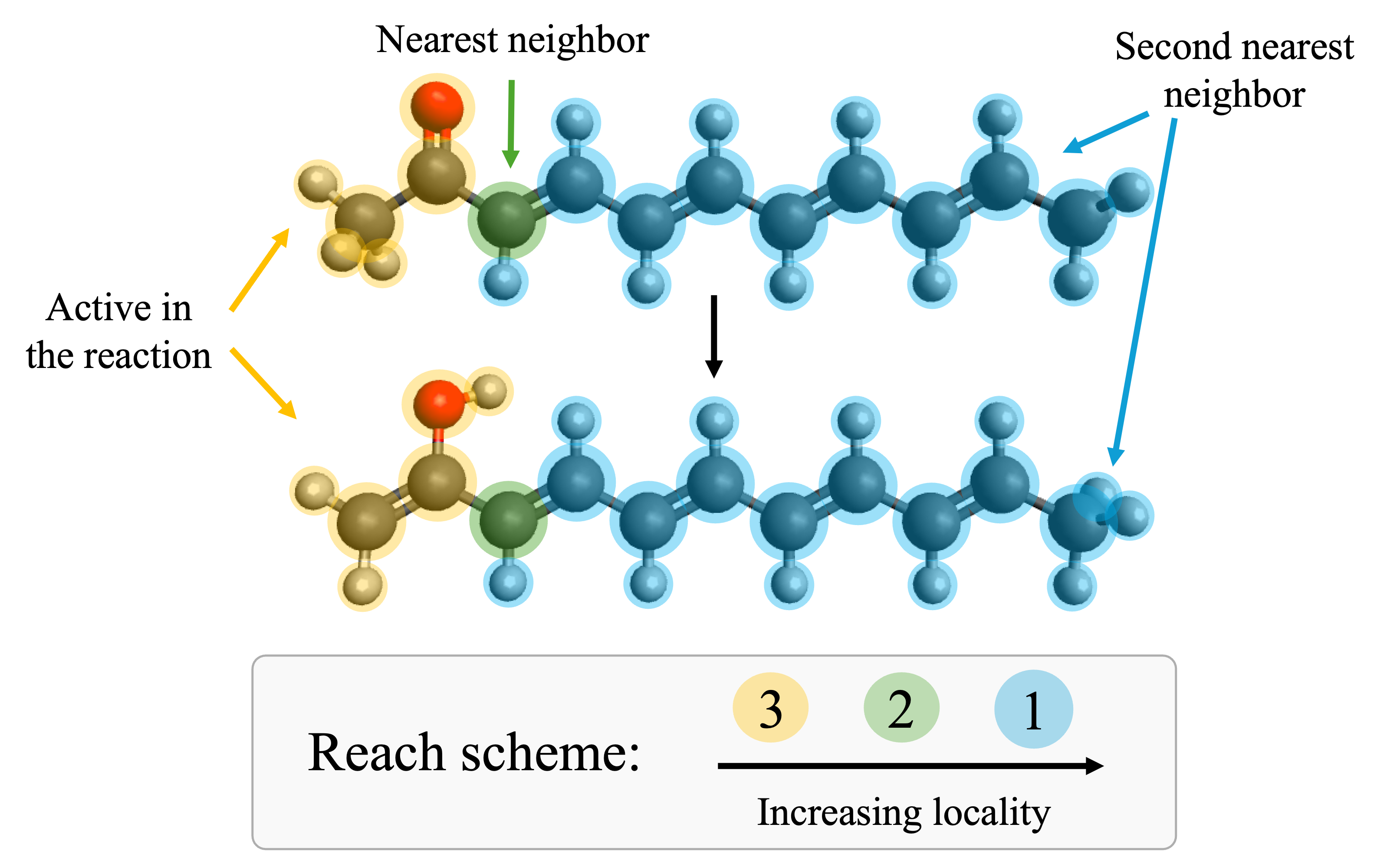}
    \caption{An example of assigning reaction-matched atomic reaches.
             In this isomerization, the atoms in yellow have had their
             bonding connectivity changed, and so they are given
             maximum reach (say, 3).
             The atoms in green are one connection removed, and so
             get a reach of one less than the maximum (e.g., 2).
             Alternatively, we could assign a minimum reach, say of
             3, in which case both the green blue atoms would
             have that reach.
    }
    \label{fig:displacement_matching}
\end{figure}

\subsection{Sparsity in the Underlying Variables}

With the machinery for solving a local subset of
the HF equations in place, we turn now to the
question of how to choose how local to make the orbitals.
We would prefer to have an approach that is simultaneously
chemically intuitive, capable of molding itself around
the needs of energy difference predictions, and
generalizable back into full HF theory via a
clear systematic limit.
To achieve these goals, we start by assuming that,
formally at least, each MO can be thought of as being
anchored on a small set of atoms.
For example, we will say that a core orbital or a lone pair
is naturally anchored on a single atom, while local
$\sigma$ bonds are naturally anchored on a pair of atoms.
What to do with less local orbitals, such as extended $\pi$
bonds, is less clear, but for the moment we take inspiration
from Mrovec and Gill's HILOs \cite{mrovecHowDelocalizedAre2024}
and allow ourselves to be guided by Lewis structures: we define
each pair of $\pi$ electrons to live in a local $\pi$
orbital that is anchored on the two atoms implied by the
molecule's Lewis structure.
In future, it may be worthwhile to explore more general
anchoring schemes that involve more than two atoms, but
for now one- and two-atom anchoring setups will suffice.
The idea is for these atoms to define the most aggressive
localization of the orbital that we will consider,
namely one in which the MO in question only incorporates
RLOs and RLVs that are also associated with one of
its anchoring atoms.
Like the MOs themselves, the RLOs and RLVs (which we define
precisely in the next section) are each anchored on one
or two atoms.
From this extremely local starting point, we will explore
how quickly accuracy is reclaimed by allowing the
MOs to spread out onto more and more neighboring atoms.\medskip

To perform this spreading systematically, we will define
the concept of an atom's reach as follows.
An atom with a reach of \textit{r} is conscious of its \textit{r} bonded connections, 
the greater consequence being the atom ignores the atoms outside of these bonded connections. 
The reach assignment becomes significant when assigning the sparsity pattern for \textbf{U} and \textbf{V}.
These assignments are made as follows, 
each MO is assigned a set of atoms: its anchor atoms and the atoms that fall 
within each anchor atoms defined reach. 
For a given MO, the elements of \textbf{U} and \textbf{V} are turned ``on'' 
for all occupied and virtual orbitals whose own anchor atoms also belong 
in this set of atoms, along with the corresponding equations. 
Note that, in $\mathbf{U}$, we always also enable
the reciprocal element, that is to say the one
that gives the MO seeded by that RLO access to
the RLO that seeded the current MO.

Now, how to set these reaches?
In this study, we test two approaches.
First, we explore the simple approach of giving every
atom the same reach.
As we will see in the results, even the modest case
of a reach of 2 proves effective at producing RHF
and MP2 isomerization energies within a kcal/mol
of the canonical theory.
As intended, we find that very large uniform reaches,
such as 12, reproduce the canonical results to a
precision limited only by the tightness of the
Cholesky threshold used in the Fock builds.
That said, we are much more interested in the behavior
of the theory in the more localized limit of small
reaches, and so test uniform reaches of 1, 2, and 3
in the results.

Second, we make a preliminary exploration of the idea
of setting an atom's reach in order to achieve a type of
\textit{reaction matching} approximation
in which orbitals unlikely to impact the reaction's
energy difference are heavily approximated while those
that are more directly involved are treated more carefully.
The basic idea is similar to excitation matching.
\cite{cluneExcitationMatchedLocal2025}
While there are a wide range of ways this idea might be
implemented, we test it in this study for isomerization
reactions by assigning a maximum reach (say, 4) to
any atom who has had its bonding connectivity changed by
the reaction, and then setting the reach for any other
atom as this maximum minus the minimum number of bonded
connections that must be traversed to get from that atom
to an atom that has maximum reach.
Figure \ref{fig:displacement_matching} shows an example.
The idea is to allow the orbitals nearby the reaction
center to be less constrained by our enforced localization,
but to keep an aggressively local
approximation in place elsewhere in the molecule to
avoid wasting effort perfecting the description
of electron pairs that are not really changing what
they are doing during the reaction and so whose
details, even if finely resolved, are unimportant
for the energy difference that we are trying to predict.

\begin{figure}[]
    \centering
    \includegraphics[width=0.95\linewidth]{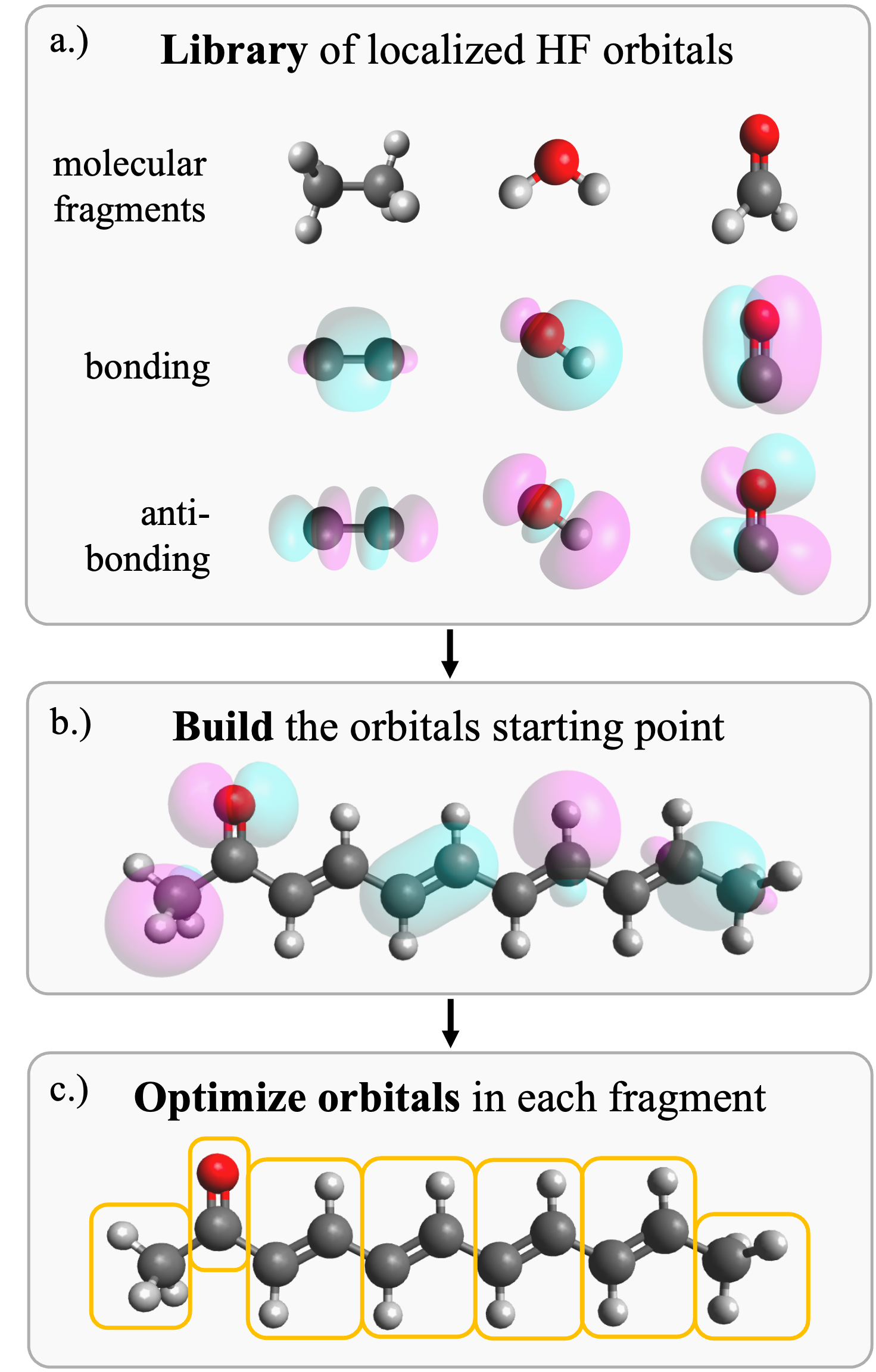}
    \caption{
    Preparation of the RLOs and the valence RLVs.
    a.) Small molecules used to construct the library of  
    Pipek-Mezey localized~\cite{pipekFastIntrinsicLocalization1989,
    lehtolaPipekMezeyOrbital2014} HF occupied/bonding 
    orbitals and virtual/anti-bonding orbitals.
    b.) Examples of library guess orbitals placed within a larger molecule.
    c.) Fragments within which small Fock diagonalizations are used
        to refine the crude initial guess.
    }
    \label{fig:flowchart}
\end{figure}
\subsection{Rough Local Orbitals}
\label{sec:RLOs}

In this study, we take a fragmentation approach to forming
the RLO and RLV orbitals.
The basic idea is to initialize highly local orbitals for
each core, local $\sigma$ bond, local $\pi$ bond, local
lone pair, local $\sigma^*$ antibond, and local $\pi^*$
antibond that one would expect a strong general chemistry
undergraduate student to be able to anticipate, as well
as a set of higher energy above-the-valence virtual
orbitals that are as local as they can be while also
being orthogonal to the occupied orbitals in their vicinity.
Of course, there are many possible ways to go about
getting a set of orbitals fitting this description.
Here, we prioritize efficiency and focus on tests involving
organic molecules formed from carbon, oxygen, and hydrogen atoms.
To this end, we have put together a simple library of stored
core, lone pair, bonding, and antibonding orbitals for a series of small
molecules: formaldehyde, methane, ethane, water, dimethyl ether, ethenol, acetaldehyde, and acetone.
Each of these stored orbitals has been truncated so that its
LCAO coefficients sit on one (for core and lone pair orbitals)
or two (for bonds and antibonds) atoms.
At the beginning of each of our HF calculations,
copies of these crude guesses for the core and valence MOs are pasted
into the actual molecule being studied, suitably rotated in 3D space
to match the local geometry.
This operation is linear scaling and has a trivial cost compared to
the later steps in the setup.
Crucially, it avoids the evaluation of two-electron integrals entirely,
allowing us to only evaluate those integrals that are needed for the
block-sparse Choleskey decomposition.

With these crude initial core and valence orbitals in place, we next set
about refining them for their local environment in the actual molecule.
To this end, we seed the $\mathbf{A}$ matrix with the crude initial
occupied orbitals and perform one whole-molecule Fock build using the
approach in Section \ref{sec:fock_build}.
With the resulting Fock matrix in hand, we proceed to carry out
a series of local fragment Roothaan-style updates to locally refine
the MOs.
We build fragments by first grouping each hydrogen atom with
whichever atom it is bonded to to create minimal fragments,
each of which is labeled by its heavy atom.
We then use the molecule's Lewis structure to pair up
any minimal fragments that share a double bond, leaving us with
final fragments that have one or two non-hydrogen atoms each,
along with the attendant hydrogens, as shown in
Figure \ref{fig:flowchart}.
We then form a small basis for each of these final fragments by
collecting together all atomic orbitals from the atoms
involved as well as any $\sigma$ bonding orbitals
at the fragment's edges.
The Fock matrix is projected into the span of these orbitals and
diagonalized, after which the resulting occupied orbitals are
localized via a modified Pipek-Mezey scheme.
\cite{cluneExcitationMatchedLocal2025}
At this point, we replace each crude initial guess occupied orbital
that lies wholly inside the fragment with its
maximally overlapping partner from the new local occupied orbitals.
These occupieds are used to initialize the set of RLOs.
After repeating for all fragments, however, we have yet to
place refined shapes for the inter-fragment $\sigma$ bonds
into the RLOs.
For each of those, we repeat the process, but using a new fragment
defined as the non-hydrogen atoms on either end of the $\sigma$
bond. 
Each of these final fragment calculations adds only the updated
$\sigma$ bond to the RLO set, again chosen by maximum overlap
with the crude initial guess for that bond.
At this point, our set of (not all orthogonal to each other)
RLO orbitals is complete.

To form the RLV orbitals, we consider our starting virtual
orbitals one at a time.
Each of these is either one of our crude guesses
for a valence anti-bonding orbital, or an individual
above-valence AO.
For each such orbital, we take the subset of RLOs
that have a population above 0.05 on its corresponding
atom or pair
of atoms, orthonormalize this subset, and project
it out of the soon-to-be-RLV.
To keep the RLVs from being too spread out, we then
zero out any LCAO coefficients that the orbital has on
atoms beyond the reach of its original atom or atom pair.
Finally, we normalize the orbital and add it to the
set of RLVs, producing orbitals that are \textit{almost}
orthogonal to the RLOs.
So long as the full-system local HF calculation does
not produce occupied orbitals that are drastically
different than the RLOs, these RLVs will satisfy the method's
requirement that they be linearly independent from the
final occupieds.
Thus, at the cost of one Fock build and a strictly linear
amount of fragment work,
we arrive at the RLO and RLV orbitals we need.

Of course, this scheme is only one way to go about producing
orbitals that satisfy the requirements that the overall method
places on the RLOs and RLVs and, at present, it is only well
defined for organic molecules with happy Lewis structures.
This setup is sufficient for the testing purposes of this
study, but, looking forward, a more general appraoch is
clearly desirable.
One set of possibilities is to adapt well-developed
fragmentation schemes, such as 
the extremely local molecular orbital (ELMO) method and its library of 
fragments\cite{sironiExtremelyLocalizedMolecular2007,meyerLibrariesExtremelyLocalized2016} or the 
fragment molecular orbital (FMO) method\cite{kitauraFragmentMolecularOrbital1999}.

\subsection{Perturbative Corrections}

Once our SCF iterations have converged, we have in hand
a set of local, non-orthogonal orbitals that approximate
localized HF orbitals.
The larger we make the atomic reaches, the smaller the
approximation, and the closer the Slater determinant
will be to the true HF Slater determinant.
At this point, we have choices to make in how we
evaluate our approximation to the HF energy.
One option is to acknowledge the fact that our orbitals
are not orthonormal, and so proceed in a cautious and
variational manner by plugging them into
$\bra{\Psi}\hat{H}\ket{\Psi}/\langle\Psi|\Psi\rangle$.
Another option is to throw caution to the wind, assume
our orbitals are close enough to orthonormal, and just
plug them into the standard HF energy formula.
A third option is to employ perturbation theory in the
singles space to correct our energy towards the true HF
energy.
In practice, we take this third approach, deriving a
singles correction formula that ends up looking a lot
like MP2, but with singles instead of doubles.

To realize this singles correction, and to follow it up
with an MP2-style doubles correction later on, it is convenient
at this stage to move to orthonormal orbitals,
which we achieve via a one-shot L\"{o}wdin orthonormalization\cite{lowdinOnTheNon1950}
of the occupied orbitals that come out of our SCF procedure.
Thus, although we avoided $O(n^3)$ cost diagonalizations
during our SCF, we permit ourselves three of them for our
singles correction, giving it a low-prefactor, non-iterative
$O(n^3)$ cost.
In small and medium molecules, this cost is trivial compared
to the Fock builds, and so we do not worry about it.
Of course, in larger systems, it would be important to modify
the singles correction to remove the $O(n^3)$ steps,
which could for example be done using techniques adopted from
local correlation methods for MP2.
For now, we simply accept these low-prefactor $O(n^3)$ steps, and
indeed our results show that they do not prevent the approach
from being highly cost competitive overall.

So, to get a perturbative singles correction, we start by
diagonalizing the occupied MO overlap matrix
$\mathbf{A}^\top\mathbf{S}\mathbf{A}$ and using the
resulting eigenvalues and eigenvectors to
L\"{o}wdin-orthonormalize our occupied orbitals.
We then perform one Fock build with these occupieds,
which, although less local now due to the orthonormalization,
are still local enough for the Cholesky-based approach to be
effective.
In our tests, this single Fock build with less-local orbitals
has a cost that is about one third of
the net cost of the Fock builds performed during the SCF procedure.
(Although our current implementation does not, one could
lower this cost further by evaluating a 
$\Delta$ Fock build that uses the difference between the final SCF
orbitals and the orthonormalized orbitals, leading to a more
aggressive screening out of the block sparse pieces of $R$.)
This new Fock matrix is then used to produce approximations
to the canonical HF occupied orbitals
via a diagonalization within the span of the occupied orbitals
that came out of our SCF.
Similarly, approximations to the canonical virtuals
are formed by first projecting these occupieds out
of the RLVs, and then diagonalizing the new Fock matrix
in the resulting basis for the virtual orbitals.
At this point, we have approximate canonical orbitals
as well as approximate occupied and
virtual orbital energies $\epsilon_i$ and $\epsilon_a$,
respectively, and we can form the occupied-virtual block
$F_{ia}^{(C)}$ of the new Fock matrix in this
approximately canonical basis.
This block is not zero, of course, because we have not
solved the full set of HF equations, but it and the
approximate orbital energies are exactly what we need
to make a second order perturbation theory correction
that shrinks the gap with standard HF theory.

To do so, we recognize that configuration interaction singles (CIS)
will offer an improved ground state description compared to
our approximate-canonical-occupieds-based final Slater determinant
$\ket{\Psi_f}$.
The approach is to start with the variational expression for
the CIS energy of our ground state
\begin{align}
    \label{eqn:E_CIS}
    E_{CIS} = \frac{\Big(\bra{\Psi_f} + \sum_{ia} t_i^a \bra{\Psi_i^a}\Big)
                    \hat{H}
                    \Big(\ket{\Psi_f} + \sum_{ia} t_i^a \ket{\Psi_i^a}\Big)}
                   {1 + \sum_{ia} (t_i^a)^2}
\end{align}
and to drop all terms that have more than two powers of the
things that we consider small.
Specifically, we will define as small for these purposes
the parts of the Hamiltonian not contained in the Fock operator
and the singles coefficients $t_i^a$ themselves.
If one then sets the derivatives of the resulting quadratic energy
expression with respect to $t_i^a$ equal to zero, solves for $t_i^a$,
and plugs those values back into the quadratic energy expression,
one arrives at a delightfully simple expression for the
singles-corrected energy.
\begin{align}
    \label{eqn:E_CIS_2}
    E_{CIS}^{(2)} = \bra{\Psi_f} \hat{H} \ket{\Psi_f}
                    + \sum_{ia} \frac{\left|F^{(C)}_{ia}\right|^2}
                                     {\epsilon_i - \epsilon_a}
\end{align}
Note that, in this formula, $i$ and $a$ sum over spin orbitals.
In our results below, the approximate HF numbers we report are
these $E_{CIS}^{(2)}$ numbers.
Again, we emphasize that, computationally, the only steps required to
get $E_{CIS}^{(2)}$ after our SCF are one additional Fock build, 
three $O(n^3)$ diagonalizations
(the occ-occ overlap, the occ-occ Fock block, and the vir-vir Fock block),
and the $O(n^2)$ sum in Eq.\ (\ref{eqn:E_CIS_2}).

In addition to using our approximate canonical orbitals in
a singles correction to improve our estimate of the HF energy
itself, we also test how well they perform when used to
estimate correlation effects via MP2 theory.
To do so, we take the straightforward approach of assuming
that our approximate canonical orbitals are in fact the
true canonical orbitals, and simply plug them into the
standard MP2 equations.
In principle, one could use the perturbative CIS correction to
the wave function to further improve our canonical orbital
approximations prior to sending them into MP2,
but we have not done so in this study.
Thus, the same approximate canonical orbitals that are used
for Eqs.\ (\ref{eqn:E_CIS}) and (\ref{eqn:E_CIS_2}) are also
used in our MP2 calculations.

\section{Results}
\begin{figure}[]
    \centering
    \includegraphics[width=0.95\linewidth]{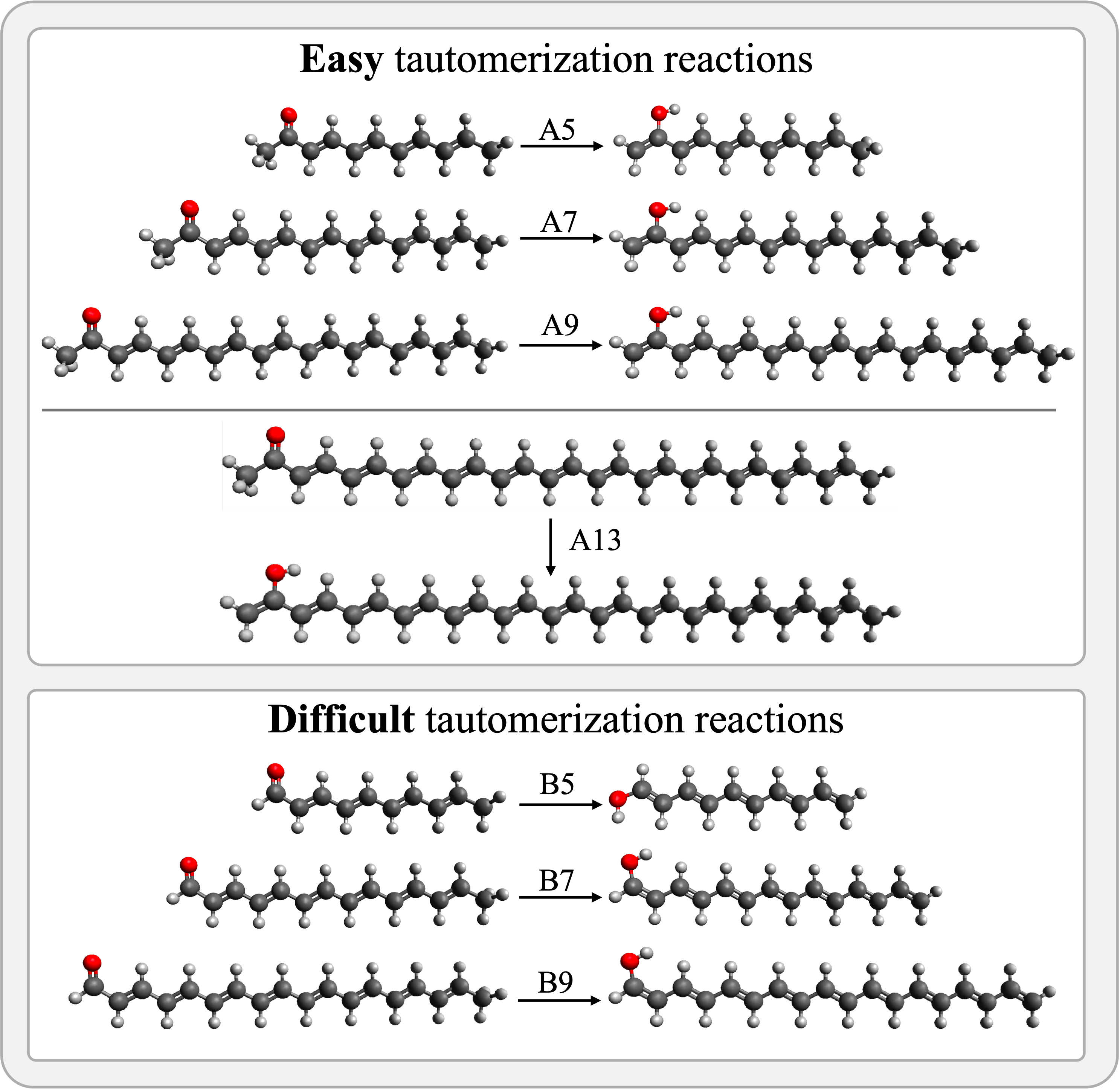}
    \caption{
    Two sets, A and B, of keto-enol tautomerization reactions on 
    fully conjugated carbon chains of increasing length to evaluate the local HF energies on.
    }
    \label{fig:molecules}
\end{figure}

\begin{figure*}[]
    \centering
    \includegraphics[width=0.95\linewidth]{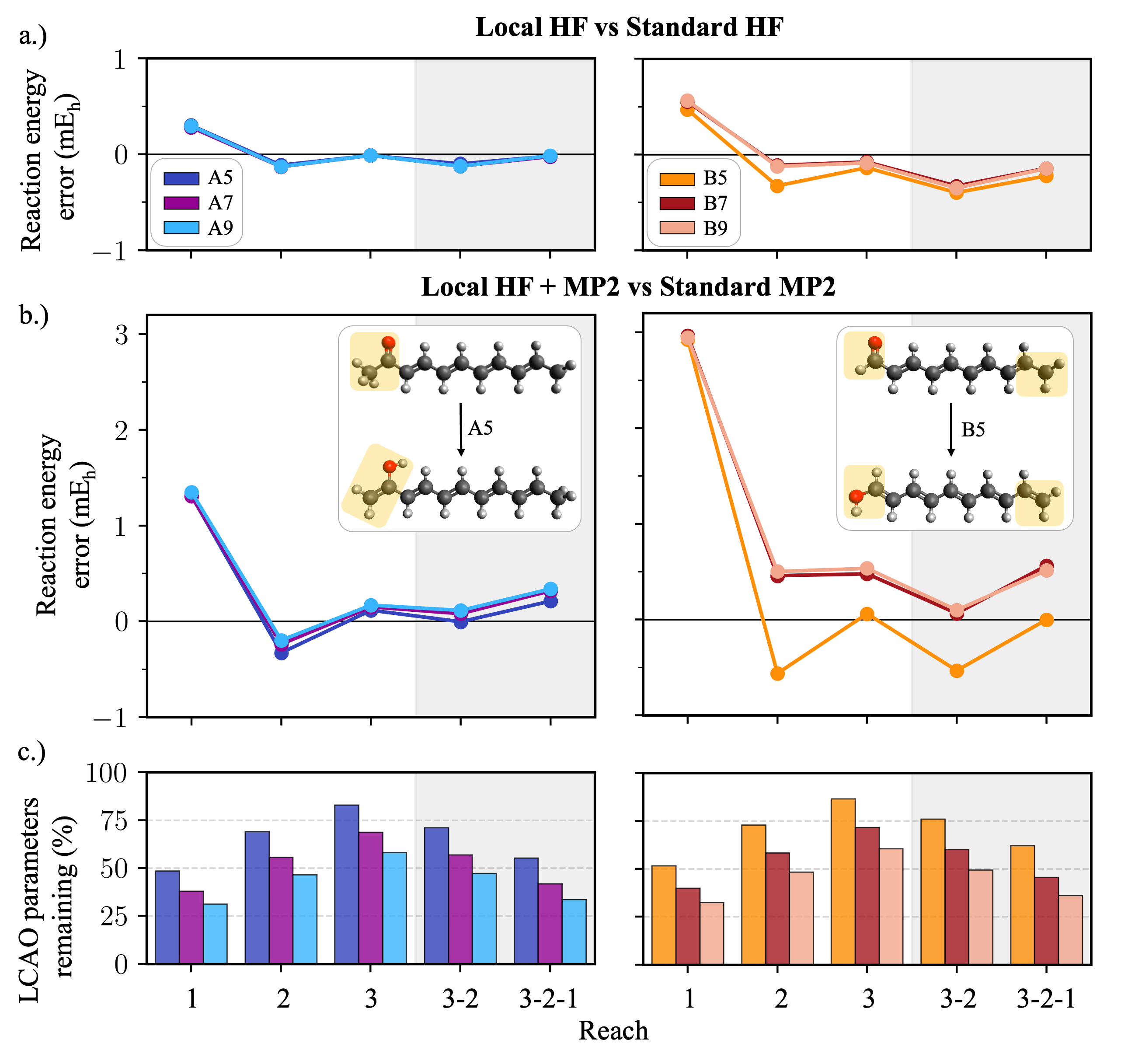}
    \caption{
    %
    Reaction energy error of a.) local HF versus standard HF, 
    b.) MP2 calculated atop local HF orbitals versus standard MP2, and 
    c.) the percent of LCAO coefficients used to construct the local HF 
    molecular orbital basis for each reaction at the different reach schemes. 
    Left: easy set of tautomerization reactions.
    Right: difficult set of tautomerization reactions.
    Unshaded: constant reach. Shaded: reaction-matched reach.
    }
    \label{fig:dE_error}
\end{figure*}
\begin{figure}
    \centering
    \includegraphics[width=0.95\linewidth]{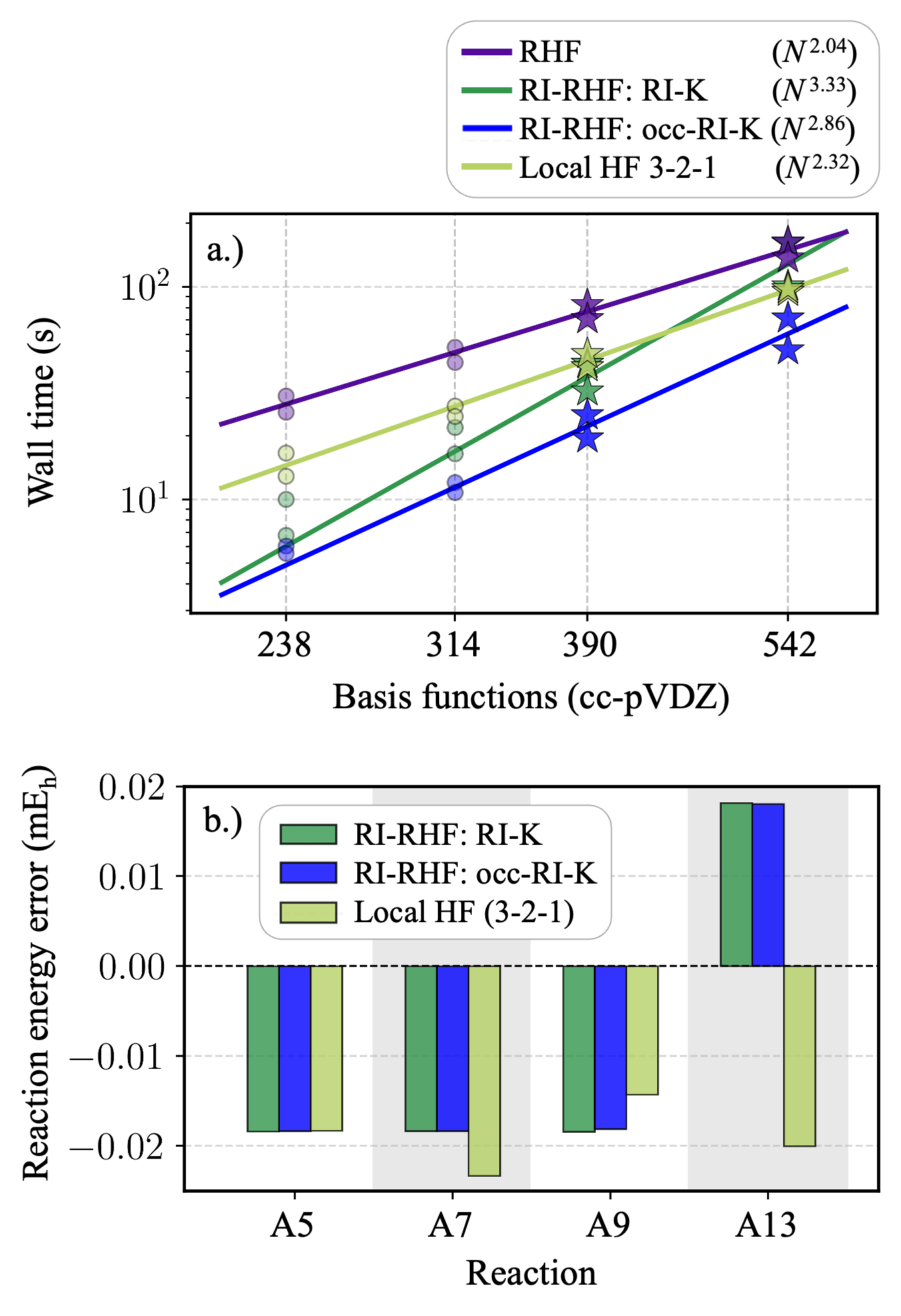}
    \caption{.
    a.) Log-Log plot of wall times of \textsc{Q-chem} RHF, \textsc{Q-chem} RI-JK RHF, 
    and our local HF across the A5/7/9/13 series of
    isomerizations.
    Note that each isomers' timing is reported as a separate point.
    Scaling slopes were found by linear fit to the largest two reactions,
    whose points are marked with stars.
    b.) The reaction energy error of RI-JK and our local approach
    vs standard HF.
    }
    \label{fig:timing}
\end{figure}
\begin{figure*}[]
    \centering
    \includegraphics[width=0.95\linewidth]{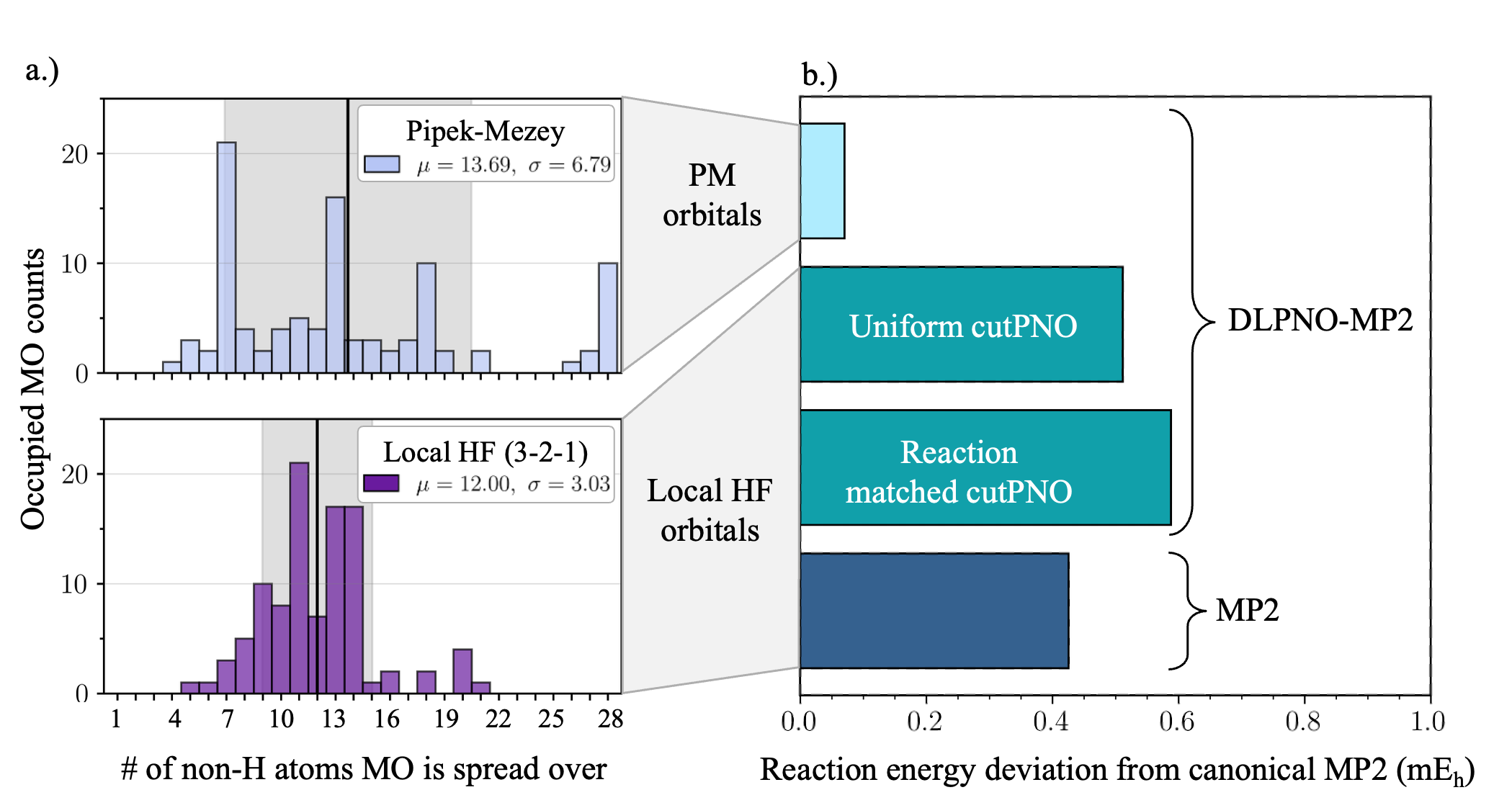}
    \caption{
    a.) Histogram of the spread of each occupied MO under PM localization and 
    local HF with a 3-2-1 reach, evaluated for the ketone tautomer of the A13 
    reaction. An MO was considered to be spread over a non-hydrogen atom if it 
    had a population greater than $10^{-8}$. \\
    b.) Comparison of the A13 reaction energy's deviation from canonical MP2 
    across several calculation setups. From top to bottom: 
    DLPNO-MP2 atop PM localized HF orbitals, 
    DLPNO-MP2 atop our 3-2-1 reach orbitals with a constant cutPNO threshold,
    DLPNO-MP2 atop our 3-2-1 reach orbitals with reaction matched cutPNO thresholds,
    and full MP2 atop our 3-2-1 reach orbitals.
    }
    \label{fig:dlpno}
\end{figure*}

\subsection{Computational details} 
\label{sec:comp_detail}

Reference RHF and MP2 energies were calculated 
in \textsc{PySCF}\cite{sunRecentDevelopmentsPySCF2020} with default settings. All 
calculations employ a cc-pVDZ basis. Local HF calculations 
used a Cholesky decomposition threshold of $1.0\times
10^{-5}$~E\textsubscript{h} and the default $\Delta E$ convergence 
threshold $1.0\times10^{-5}$~E\textsubscript{h}. 
Molecular geometries were optimized at the MP2/cc-pVDZ 
level of theory and are available in the SI. The frozen 
core approximation was not applied in any of the results.

To test the local HF method across reactions of varying enforced 
locality, we consider two sets of keto-enol tautomerization 
reactions on fully conjugated carbon chains of increasing 
length (see Fig.~\ref{fig:molecules}). In set A, the 
structural change between isomers is 
confined to the ketone and enol groups at one end of the 
molecule, making the reaction energy relatively local in 
character. Set B is constructed analogously but with the 
terminal methyl group removed, shifting the 
$\pi$-system by one bond between isomers and introducing 
bond changes at both ends of the molecule. This makes set B 
a more difficult test of our local approach, as its
isomerization changes the $\pi$ conjugation pattern across
the entire molecule.
Set A contains four reactions whose 
conjugated chains contain five, seven, nine, and 
thirteen $\pi$ bonds, referred to as A5/A7/A9/A13, and 
set B contains the corresponding first three reactions, 
B5/B7/B9, respectively. 

All DLPNO-MP2 calculations were performed in an in-house 
code. DLPNO calculations using PM orbitals 
and the 3-2-1 reach orbitals under a uniform
cutPNO threshod set this
threshold to $10^{-7}$.
In the reaction-matched approach to setting cutPNO,
it was set to values of
$10^{-7}$, $10^{-5}$, or $10^{-3}$ for
$ij$ pairs for which the maximum-reach atoms within
their outer range had reach 3, 2, or 1, respectively.
For this purpose, the outer range was defined as all
atoms on which either of the orbitals has a population
of at least 0.001, as well as each atom that
has a projected atomic orbital with an LCAO coefficient
above 0.1 on any of those atoms.
All calculations increased the cutPNO threshold by a factor of 
$1\times10^{-2}$ for any orbital pair with a core orbital, 
as is the standard when not using the frozen core approximation.\cite{noauthor_713_nodate}

\subsection{Easy isomerization reactions with uniform reaches}

As an initial demonstration of the local HF method's 
accuracy, we analyze the reaction energy errors for 
set A, shown in the left column of Fig.~\ref{fig:dE_error}. 
Focusing first on the unshaded region, in which we plot
results for the case of all atoms being given the same reach,
we see an overall systematic improvement 
in the error with increasing reach in both the HF and MP2 
results.
With increasing reach, the constraints in the local 
HF solver are loosened, allowing more basis functions to 
contribute to each molecular orbital. In the limit of 
complete LCAO flexibility, the true HF result is recovered, 
so this convergent behavior is to be expected.
The HF reaction energy error in 
Fig.~\ref{fig:dE_error}a remains below $0.5$~mE\textsubscript{h} 
across all reaches, well within chemical accuracy of 
$1.6$~mE\textsubscript{h}. It is worth noting that, while 
this sub-mE\textsubscript{h} accuracy in the reaction energy 
is achieved at all reaches, the absolute energy error for 
an individual molecule remains on the order of tens of 
mE\textsubscript{h}, confirming that the local HF 
approximation using uniform reaches achieves effective error
cancellation between isomers.

Turning to the MP2 results in Fig.~\ref{fig:dE_error}b, 
the reaction energy errors vs standard MP2 are somewhat larger
than what we see for the HF energy errors.
That said, even a uniform reach of one achieves reaction
energies within $1.6$~mE\textsubscript{h} of standard MP2,
and the reach two and reach three results are all
within $0.5$~mE\textsubscript{h}.
reaction energy to sub-mE\textsubscript{h} accuracy while 
maintaining a sparse LCAO coefficient matrix, as 
shown in Fig.~\ref{fig:dE_error}c.

\subsection{Difficult isomerization reactions with uniform reaches}

Having demonstrated that local HF performs well for a highly
localized isomerization, we now turn to set B, 
the more difficult keto-enol tautomerization shown in the right column of 
Fig.~\ref{fig:dE_error}. Because the whole $\pi$-system shifts 
by one bond between isomers, it is much less obvious that a local
theory should be effective.
Looking at the unshaded region of Figs.\ \ref{fig:dE_error}a
and \ref{fig:dE_error}b,
we see that the errors are indeed larger than in the easier
reaction, but the same overall trends hold.
The reach two and reach three results are significant improvements on
the reach one results, and, starting at reach two, the reaction
energies are within about $0.5$~mE\textsubscript{h} of those of
the canonical theories.
Thus, despite the presence of an extended $\pi$ system that
shifts its conjugation in response to a reaction that moves a hydrogen
atom from one end of the molecule to the other, our local approach
performs well under remarkably modest uniform reach schemes.
Looking at Fig.~\ref{fig:dE_error}c, we see that, in the largest
systems, the reach two scheme achieves its accuracy despite disabling
more than half of the underlying orbital variables.
We also see that, as expected, the fraction of the full variable set
that gets used decreases as we go to larger and larger molecules.

\subsection{Isomerization reactions with reaction matched reaches}

So far, we have analyzed the results when all atoms are
assigned a uniform reach, but one of the key questions we have
is whether this study's approach to HF theory allows its
approximation to be more directly tailored to the reaction
in question.
To test this idea, we have also carried out calculations
in which the atomic reaches are set based on proximity
to the reactive atoms (which are those highlighted in yellow
in Fig.~\ref{fig:dE_error}).
Specifically, we present two tests.
In the 3-2 reach test, the reactive atoms are each assigned
a reach of 3, while the rest get a reach of 2.
In the 3-2-1 reach test, the reactive atoms are each assigned
a reach of 3, their nearest neighbors a reach of 2, and
all other atoms a reach of just 1.
As seen in Fig.~\ref{fig:dE_error}c, these approaches
substantially increase the fraction of the underlying
variables that get set aside and neglected during the SCF procedure.

Looking at the shaded regions of 
Fig.~\ref{fig:dE_error}a and Fig.~\ref{fig:dE_error}b,
we see that both of these tailored reach schemes achieve
reaction energies within about $0.5$~mE\textsubscript{h}
of canonical HF and MP2.
This result confirms that, as expected, most MOs can have
aggressive locality enforcement without damaging the
energy difference so long as those MOs most
strongly involved in the reaction are given a little bit
more freedom.
Thanks to our reorganization of the HF equations into a
form in which it is easy to map a given degree of locality
for a specific MO onto the disabling of a well defined
set of variables and corresponding equations, these types
of systematic reach schemes are straightforward to carry out.

\subsection{HF timing comparison}

Having shown that the local approach is capable of accurately
approximating canonical HF and MP2 reaction energies,
we now turn our attention to the question of how effective
it is at reducing costs.
Such comparisons are not straightforward, in particular
because the long study of HF theory has produced highly
optimized code bases, with a quality of underlying software
engineering that our our initial implementation of the
local approach should not pretend to be comparable to.
That said, we are nonetheless able to offer preliminary
evidence that the hoped for cost savings are real and that
they appear even in molecules of modest size.
In other words, the approach works to reduce cost long before
one reaches the large system limit.

For quantitative timing comparisons, we have tested
our method against three well-established HF implementations
in the A5/A7/A9/A13 isomerization series,
namely Q-Chem's  \cite{epifanovskySoftwareFrontiersQuantum2021}
standard RHF implementation and two of its 
resolution-of-the-identity methods: the
older RI-JK\cite{weigend_fully_2002} and the more
recent occ-RI-K\cite{manzer2015fast} 
exchange algorithms.
For our approach, the timings include integral
generation (performed using Libint\cite{valeyevEvaleevLibint2702021}),
the construction of the Cholesky factors, the setup of the RLOs
and RLVs, the SCF iterations,
and the final $E_{CIS}^{(2)}$ correction.
Among these, we find that the construction of the Cholesky factors
is the leading cost, followed by the Fock builds used within the SCF
and the perturbative correction.
The GMRES solver and the setup of the RLOs and RLVs take very
little time in comparison.
For Q-Chem, the timings include integral generation,
initial guess generation via Q-Chem's default guess,
and the SCF iterations.
For clarity of comparison, all timing calculations were run
using a single thread on the same computational node, which was
equipped with an Intel Xeon Gold 6330 processor.
To ensure as fair a comparison as possible,
the energetic convergence tolerances for all methods were set to
$10^{-6}$ E\textsubscript{h}.
For Q-Chem, the two-electron integral screening threshold 
was set to $10^{-10}$, while our
Cholesky decomposition threshold was set to $10^{-5}$
E\textsubscript{h} (at which level our method produces similar
HF reaction energy errors as the RI-JK method, see below).

The timing results are 
displayed in a log-log plot in Figure \ref{fig:timing}a.
As expected at these modest system sizes, the standard HF
implementation is dominated by the $O(n^2)$ cost of screened
two-electron integral evaluation, with the $O(n^3)$ term
from the Roothaan diagonalization not yet in evidence.
Remarkably, the local approach is more efficient than this
standard HF implementation across the full series and
also displays a cost scaling dominated by $O(n^2)$ terms.
The $O(n^3)$ terms are somewhat more in evidence,
which is likely because our implementation does not exploit the
sparsity in the $\tilde{\mathbf{A}}$, $\tilde{\mathbf{B}}$,
$\mathbf{U}$, and $\mathbf{V}$ matrices when carrying out
its GMRES solves, and the resulting pile of $O(n^3)$ operations
is not as numerically efficient as
the symmetric diagonalization LAPACK routine at the heart
of the $O(n^3)$ Roothaan solver.
Nonetheless, the observed cost scaling of the local approach is
lower than the RI methods, which are both expected to limit to
$O(n^4)$ scaling in larger systems\cite{
weigandApproximated2009,dunlapRobustVariationalFitting2000} 
(interestingly, these system sizes are small enough that
sub-dominant scaling terms are more important for occ-RI-K).
Although our timings are not quite as efficient as the newer RI method,
they do manage an explicit crossover with the older RI method
at around 400 basis functions, which we consider especially
significant given Cholesky methods' historical cost disadvantages
vs RI.\cite{weigandApproximated2009} 
As seen in Figure \ref{fig:timing}b, the accuracy vs standard
HF of the 3-2-1 reach scheme used in this timing test is
comparable to that of the RI methods, confirming that, at the HF
level at least, the observed cost competitiveness is achieved
in an accuracy-balanced setting.
All that said, it is crucial to emphasize the preliminary nature
of these results.
These RI methods are general purpose HF methods that can be used in a
much wider variety of settings than this study's initial
setup of the local approach, and further work will be needed
to determine how timings compare once
the local approach is generalized beyond
simple organic molecules near their equilibrium geometries
and adapted to use RI integral factorization.

\subsection{Local correlation}

Beyond HF itself, more local orbitals should also
be useful in local correlation approaches, and so
we also perform some preliminary tests on the
accuracy consequences of using our orbitals and
a reaction matching scheme in domain localized
pair natural orbital MP2 (DLPNO-MP2).\cite{pinskiSparseMapsSystematic2015,
riplingerEfficientLinearScaling2013}
To start, it is worth inspecting how much more local
our imposed-locality orbitals are compared to those from a unitary remixing
approach.
To this end, Fig.\ \ref{fig:dlpno}a shows
distributions of MO locality in the A13 reaction's
ketone isomer, both for standard HF orbitals under
Pipek-Mezey localization and for our 3-2-1 reach scheme.
The 3-2-1 orbitals are on average spread over fewer
atoms than the Pipek-Mezey orbitals, and the least
local of them are notably more local than the 
least local Pipek-Mezey orbitals.
Thus, as intended, the approach succeeds in producing
orbitals that are more local than what can be
achieved without approximating the HF wave function.

In Fig.\ \ref{fig:dlpno}b, we see that the
deviation in A13 from the standard MP2 reaction energy
that is created by using the 3-2-1 reach orbitals
in DLPNO-MP2 is only slightly larger than that
seen when using them in standard MP2, and that both
of these deviations are well below the 1 kcal/mol
(1.6 mE\textsubscript{h}) level.
When the reaction matching approach was also used
to set varying cutPNO thresholds
(see Section \ref{sec:comp_detail}),
the deviation increases slightly, but again is
well within 1 kcal/mol of the canonical MP2 prediction.
Thus, although this demonstration is far from
a full analysis of these extra-local orbitals'
ability to support local correlation methods,
it does suggest that the errors produced by this
pairing are likely to be similar in size as those
produced by pairing these orbitals with
canonical correlation methods.

\section{Conclusion}

We have investigated a reformulation of the HF equations intended
to facilitate strong locality constraints on the molecular orbitals.
We find that, as desired, this formulation allows large fractions
of the LCAO coefficients to be disabled while retaining a low-overhead
SCF solver.
In tests on isomerization reactions in linear conjugated molecules,
we find that the approach is highly cost competitive, outperforming
a production-level implementation of the standard SCF algorithm and
competing with resolution-of-the-identity methods even in molecules
of very modest size.
Despite using locality constraints to disable roughly half of the
orbitals' variational flexibility, we find that predicted HF and
MP2 reaction energies are close to their canonical values, even
when spatially heterogeneous locality constraints are used.
These results provide strong preliminary evidence that it is
possible to benefit computationally from imposing locality on the
HF orbitals well before one reaches the large molecule regime.

Looking forward, there is much to do.
Obviously, the method used here to prepare the rough local orbitals
is far from general.
Possible routes to a more general approach include leveraging
existing fragmentation methodology that exists for SCF initial
guesses, as well as the use of lower-level theories such as
DFT-B and other semi-empirical methods for the generation of
an initial orbital set.
Similarly, the Cholesky approach used here to exploit orbital
locality during the Fock build is only one such approach,
and it is unlikely to be the fastest option.
RI-based approaches should also be tested.
Finally, we have not yet exploited the sparsity present
within the inner GMRES solver, and this shortcoming will be
increasingly noticeable as other parts of the methodology
are made faster.
Like the Fock build approach, it should be possible to organize
the GMRES operations in a block-sparse fashion in order to
simultaneously achieve lower scaling while maintaining the
efficiency of dense matrix multiplication.
Finally, it will be interesting to explore more thoroughly
the effect that the extra-local orbitals enabled by this
approach have on local correlation methods in a wider variety of settings.

\section*{Supplementary Information}
See \href{run:SI.tex}{Supplementary Information} for...

\begin{acknowledgments}
This work was supported by the Office of Science, Office of Basic Energy Sciences, 
the U.S. Department of Energy, Contract Number DE-AC02-05CH11231, 
through the Gas Phase Chemical Physics program. Computational work was performed 
with the LBNL Lawrencium cluster. 
T.K.Q. acknowledges that this material is based upon work supported by the 
National Science Foundation Graduate Research Fellowship Program under Grant No. DGE 2146752. 
T. K. Q. acknowledges support from the Molecular Sciences Software Institute, 
funded by the National Science Foundation (CHE-2136142). We would like to thank 
Dr. Taylor Barnes for helpful discussions and advice regarding the software implementation. 
Any opinions, findings, and conclusions or recommendations expressed in this 
material are those of the authors and do not necessarily reflect the views of the 
National Science Foundation. 
\end{acknowledgments}

\section*{Data Availability Statement}

The data that support the findings of this study are available
within the article and its supplementary material.

\section*{References}
\bibliographystyle{achemso}
\bibliography{lhf}

\clearpage
\onecolumngrid
\section{Supplementary Information}
\renewcommand{\thesection}{S\arabic{section}}
\renewcommand{\theequation}{S\arabic{equation}}
\renewcommand{\thefigure}{S\arabic{figure}}
\renewcommand{\thetable}{S\arabic{table}}
\setcounter{section}{0}
\setcounter{figure}{0}
\setcounter{equation}{0}
\setcounter{table}{0}
All coordinates are in Angstrom.
%
\begin{table}[h]
\centering
\begin{tabular}{lrrr}
34 &  &  &  \\
\# A1 ketone&  &  &  \\
O & 4.81367490 & -1.37523493 & -0.20002452 \\
C & 4.79485852 & -0.15762609 & -0.07759628 \\
C & 3.48652737 & 0.62055221 & 0.00394263 \\
H & 3.48226588 & 1.36662055 & -0.81508594 \\
H & 3.49895333 & 1.21517526 & 0.93853934 \\
C & 2.24581105 & -0.26629532 & -0.06196927 \\
H & 2.28613139 & -1.00458583 & 0.75890298 \\
H & 2.27262085 & -0.85455158 & -0.99657690 \\
C & 0.94199701 & 0.53099860 & 0.01656199 \\
H & 0.92342169 & 1.11815009 & 0.95558126 \\
H & 0.90908599 & 1.26803890 & -0.80959743 \\
C & -0.30553140 & -0.35441675 & -0.04852173 \\
H & -0.27283066 & -1.09088596 & 0.77757948 \\
H & -0.28698459 & -0.94112896 & -0.98730698 \\
C & -1.61435608 & 0.43629688 & 0.02910917 \\
H & -1.63179214 & 1.02283929 & 0.96832948 \\
H & -1.64568419 & 1.17315562 & -0.79704133 \\
C & -2.86194274 & -0.44911464 & -0.03646972 \\
H & -2.83010383 & -1.18615606 & 0.78937672 \\
H & -2.84426827 & -1.03543728 & -0.97571118 \\
C & -4.17090604 & 0.34108646 & 0.04150923 \\
H & -4.18839396 & 0.92770016 & 0.98088588 \\
H & -4.20241615 & 1.07865533 & -0.78423268 \\
C & -5.41922270 & -0.54312694 & -0.02419349 \\
H & -5.38529183 & -1.27941824 & 0.80079885 \\
H & -5.39934826 & -1.12853030 & -0.96269990 \\
C & -6.71975543 & 0.25979105 & 0.05487067 \\
H & -7.60553093 & -0.39499554 & 0.00589648 \\
H & -6.77259171 & 0.83035081 & 0.99816933 \\
H & -6.78672885 & 0.98223346 & -0.77683459 \\
C & 6.07129262 & 0.66361725 & 0.00264690 \\
H & 6.94569301 & 0.00387754 & -0.08729958 \\
H & 6.11123475 & 1.20619311 & 0.96250853 \\
H & 6.08858554 & 1.42009497 & -0.80008366 \\
\end{tabular}
\end{table}

\begin{table}[h]
\centering
\begin{tabular}{lrrr}
34 &  &  &  \\
\# A1 enol&  &  &  \\
O & 5.01409000 & -1.04301900 & 0.95018000 \\
H & 4.35386300 & -0.86756400 & 1.63428400 \\
C & 4.78133700 & -0.15674300 & -0.07738500 \\
C & 3.49685300 & 0.62493500 & 0.00422700 \\
H & 3.47052700 & 1.36385700 & -0.81387800 \\
H & 3.48164600 & 1.19944100 & 0.95330800 \\
C & 2.24900700 & -0.27156100 & -0.06246500 \\
H & 2.28565300 & -1.03035000 & 0.74261300 \\
H & 2.26873000 & -0.83438700 & -1.01426900 \\
C & 0.94182800 & 0.51671200 & 0.04924000 \\
H & 0.93405800 & 1.08272100 & 1.00120300 \\
H & 0.90092700 & 1.27110100 & -0.76016200 \\
C & -0.30281100 & -0.37213300 & -0.02282200 \\
H & -0.25907800 & -1.12757700 & 0.78573800 \\
H & -0.29290600 & -0.93658200 & -0.97520300 \\
C & -1.61347500 & 0.41186400 & 0.08788000 \\
H & -1.62154400 & 0.97752300 & 1.03989500 \\
H & -1.65691000 & 1.16637900 & -0.72144500 \\
C & -2.85842500 & -0.47694400 & 0.01770400 \\
H & -2.81423500 & -1.23160300 & 0.82693500 \\
H & -2.84989400 & -1.04249100 & -0.93428600 \\
C & -4.16898900 & 0.30680000 & 0.12850900 \\
H & -4.17674000 & 0.87325000 & 1.08029900 \\
H & -4.21296300 & 1.06155900 & -0.68092000 \\
C & -5.41513900 & -0.58037600 & 0.05932700 \\
H & -5.36864000 & -1.33384900 & 0.86807400 \\
H & -5.40491100 & -1.14565700 & -0.89160200 \\
C & -6.71719300 & 0.21625400 & 0.17138000 \\
H & -7.60125900 & -0.44055200 & 0.11942700 \\
H & -6.76029200 & 0.76650400 & 1.12714000 \\
H & -6.79686200 & 0.95580700 & -0.64397400 \\
C & 5.66370600 & -0.06793700 & -1.09404400 \\
H & 5.47005600 & 0.61177400 & -1.92485100 \\
H & 6.57027400 & -0.67745900 & -1.09708600 \\
\end{tabular}
\end{table}

\begin{table}[h]
\centering
\begin{tabular}{lrrr}
32 &  &  &  \\
\# A2 ketone&  &  &  \\
O & 4.81013622 & -1.10167237 & -1.02858842 \\
C & 4.65884128 & -0.21377767 & -0.19538203 \\
C & 3.37189514 & 0.54156126 & -0.06242938 \\
H & 3.30277570 & 1.32542334 & 0.70280999 \\
C & 2.31901095 & 0.27037369 & -0.86685268 \\
H & 2.45618884 & -0.52892265 & -1.60984413 \\
C & 0.98116087 & 0.94502171 & -0.80093430 \\
H & 1.00947989 & 1.77547563 & -0.07215463 \\
H & 0.74660258 & 1.38626984 & -1.78901805 \\
C & -0.14080544 & -0.03876016 & -0.42565701 \\
H & 0.08219427 & -0.47807654 & 0.56498036 \\
H & -0.14558359 & -0.87854141 & -1.14669534 \\
C & -1.52357896 & 0.61806496 & -0.39869841 \\
H & -1.51350706 & 1.46272087 & 0.31732418 \\
H & -1.73962795 & 1.05544686 & -1.39277650 \\
C & -2.64310075 & -0.35485399 & -0.01838118 \\
H & -2.42469069 & -0.79209326 & 0.97519012 \\
H & -2.65123116 & -1.19963891 & -0.73417508 \\
C & -4.02778062 & 0.29781896 & 0.00947601 \\
H & -4.01905951 & 1.14348085 & 0.72474120 \\
H & -4.24580131 & 0.73535371 & -0.98440195 \\
C & -5.14864604 & -0.67345647 & 0.39010323 \\
H & -4.92830757 & -1.10976689 & 1.38256757 \\
H & -5.15478724 & -1.51752830 & -0.32499886 \\
C & -6.52665465 & -0.00802754 & 0.41394824 \\
H & -7.31867944 & -0.72369966 & 0.68990539 \\
H & -6.55047589 & 0.82052105 & 1.14268612 \\
H & -6.77839859 & 0.41002220 & -0.57597775 \\
C & 5.76406876 & 0.18788263 & 0.76464110 \\
H & 6.65821584 & -0.42518729 & 0.58607657 \\
H & 5.42215678 & 0.05956087 & 1.80615996 \\
H & 6.00788487 & 1.25591804 & 0.63027679 \\
\end{tabular}
\end{table}
\begin{table}[h]
\centering
\begin{tabular}{lrrr}
32 &  &  &  \\
\# A2 enol&  &  &  \\
O & 4.85380700 & -1.01360800 & -1.27667300 \\
H & 4.32866300 & -0.62988400 & -1.99466200 \\
C & 4.66131000 & -0.20287100 & -0.18294700 \\
C & 3.38454500 & 0.51944600 & -0.07882400 \\
H & 3.32757000 & 1.29270800 & 0.69756000 \\
C & 2.30389200 & 0.28158200 & -0.86289300 \\
H & 2.34738100 & -0.53115300 & -1.60353100 \\
C & 0.99165100 & 1.00480100 & -0.74778000 \\
H & 1.08651400 & 1.83451600 & -0.02351400 \\
H & 0.72974000 & 1.45894800 & -1.72334600 \\
C & -0.15163000 & 0.07062000 & -0.31805600 \\
H & 0.09847200 & -0.37457000 & 0.66335200 \\
H & -0.22370100 & -0.77120100 & -1.03370500 \\
C & -1.50449900 & 0.78251800 & -0.23309100 \\
H & -1.42890000 & 1.62667700 & 0.47959100 \\
H & -1.74589200 & 1.22792200 & -1.21782800 \\
C & -2.64557900 & -0.14307900 & 0.19792600 \\
H & -2.40173100 & -0.58910400 & 1.18157800 \\
H & -2.72031700 & -0.98684900 & -0.51529000 \\
C & -3.99957600 & 0.56620500 & 0.28571100 \\
H & -3.92449500 & 1.41013400 & 0.99908600 \\
H & -4.24261700 & 1.01335400 & -0.69805500 \\
C & -5.14206600 & -0.35801300 & 0.71596100 \\
H & -4.89660200 & -0.80423300 & 1.69803900 \\
H & -5.21479200 & -1.20032200 & 0.00243800 \\
C & -6.48886700 & 0.36407000 & 0.80000100 \\
H & -7.29711200 & -0.31836300 & 1.11091200 \\
H & -6.44614500 & 1.19193500 & 1.52862900 \\
H & -6.76628700 & 0.79355600 & -0.17810200 \\
C & 5.61911700 & -0.12861000 & 0.77177100 \\
H & 5.44109900 & 0.45651600 & 1.67585800 \\
H & 6.55723900 & -0.67637000 & 0.65986000 \\
\end{tabular}
\end{table}

\begin{table}[h]
\centering
\begin{tabular}{lrrr}
30 &  &  &  \\
\# A3 ketone&  &  &  \\
O & 4.72681982 & -1.25823698 & -0.80232254 \\
C & 4.62217889 & -0.24130480 & -0.12184925 \\
C & 3.31846238 & 0.46814728 & 0.04587005 \\
H & 3.28722290 & 1.37214545 & 0.66726371 \\
C & 2.19157792 & 0.00557843 & -0.55816369 \\
H & 2.28722501 & -0.90375606 & -1.16776009 \\
C & 0.88160528 & 0.62857430 & -0.45250637 \\
H & 0.79542788 & 1.53745013 & 0.15984728 \\
C & -0.22318722 & 0.14051944 & -1.07534759 \\
H & -0.11739035 & -0.77184191 & -1.68197498 \\
C & -1.59815885 & 0.73310565 & -0.96951553 \\
H & -1.55164753 & 1.68914129 & -0.41633356 \\
H & -1.98051234 & 0.96418425 & -1.98287131 \\
C & -2.58925589 & -0.21836203 & -0.27663220 \\
H & -2.21945993 & -0.44244791 & 0.74177917 \\
H & -2.61066864 & -1.18220225 & -0.82093134 \\
C & -4.00758184 & 0.35234656 & -0.19707185 \\
H & -3.98297290 & 1.31992212 & 0.34135901 \\
H & -4.37011064 & 0.57531388 & -1.21958856 \\
C & -4.99683511 & -0.58766471 & 0.49735031 \\
H & -4.63044881 & -0.80948005 & 1.51725855 \\
H & -5.01791849 & -1.55350061 & -0.04161979 \\
C & -6.41121977 & -0.00782919 & 0.57200716 \\
H & -7.10698807 & -0.69934057 & 1.07522875 \\
H & -6.41692500 & 0.94392881 & 1.13073707 \\
H & -6.80696601 & 0.19474031 & -0.43813032 \\
C & 5.80928662 & 0.37133659 & 0.59977469 \\
H & 6.71232930 & -0.22504829 & 0.40986655 \\
H & 5.60889021 & 0.41376870 & 1.68428203 \\
H & 5.96473477 & 1.40816350 & 0.25505923 \\
\end{tabular}
\end{table}
\begin{table}[h]
\centering
\begin{tabular}{lrrr}
30 &  &  &  \\
\# A3 enol&  &  &  \\
O & 4.73834200 & -1.21539400 & -1.06341100 \\
H & 4.12429900 & -0.97993100 & -1.77482400 \\
C & 4.62551900 & -0.22796400 & -0.11172500 \\
C & 3.33077700 & 0.44369600 & 0.03192800 \\
H & 3.31232900 & 1.33866500 & 0.66609000 \\
C & 2.17592300 & 0.01668900 & -0.55434500 \\
H & 2.18041600 & -0.90898900 & -1.14737400 \\
C & 0.89688300 & 0.69973500 & -0.42152900 \\
H & 0.87832800 & 1.61940900 & 0.18096600 \\
C & -0.25434800 & 0.26583400 & -0.99811700 \\
H & -0.22410900 & -0.65755300 & -1.59690000 \\
C & -1.59016000 & 0.93545200 & -0.84675300 \\
H & -1.46686100 & 1.89533900 & -0.31209300 \\
H & -2.00278400 & 1.17455800 & -1.84640800 \\
C & -2.60334300 & 0.05364500 & -0.09650600 \\
H & -2.20378400 & -0.17462400 & 0.90966600 \\
H & -2.70155000 & -0.91630800 & -0.62140500 \\
C & -3.98321700 & 0.70442400 & 0.03087100 \\
H & -3.88205600 & 1.67764700 & 0.54987400 \\
H & -4.37563800 & 0.93120500 & -0.97975400 \\
C & -4.99355900 & -0.16659400 & 0.78261500 \\
H & -4.59740500 & -0.39259600 & 1.79040400 \\
H & -5.09154700 & -1.13821400 & 0.26279700 \\
C & -6.36887200 & 0.49362800 & 0.90508500 \\
H & -7.08055100 & -0.14916100 & 1.44905900 \\
H & -6.29757700 & 1.45288500 & 1.44620000 \\
H & -6.79493000 & 0.70247200 & -0.09137700 \\
C & 5.68671500 & 0.05688500 & 0.68270000 \\
H & 5.58232500 & 0.78569200 & 1.48867700 \\
H & 6.63784000 & -0.46148100 & 0.54339400 \\
\end{tabular}
\end{table}

\begin{table}[h]
\centering
\begin{tabular}{lrrr}
28 &  &  &  \\
\# A4 ketone&  &  &  \\
O & 4.69398253 & -1.33996651 & -0.63778762 \\
C & 4.62619613 & -0.20112243 & -0.18244580 \\
C & 3.32239961 & 0.50481409 & -0.00905652 \\
H & 3.32422123 & 1.52188431 & 0.40324478 \\
C & 2.15161201 & -0.09860719 & -0.35502450 \\
H & 2.21586705 & -1.11732037 & -0.76241840 \\
C & 0.84207935 & 0.50650173 & -0.22305460 \\
H & 0.78396873 & 1.52541131 & 0.18510831 \\
C & -0.31216767 & -0.13240098 & -0.58191571 \\
H & -0.24387820 & -1.15139853 & -0.98970941 \\
C & -1.63714707 & 0.45172975 & -0.45989480 \\
H & -1.70365704 & 1.47072776 & -0.05188973 \\
C & -2.77932702 & -0.18972001 & -0.82412860 \\
H & -2.69805005 & -1.21019413 & -1.22897408 \\
C & -4.16171045 & 0.37804593 & -0.68187479 \\
H & -4.10042664 & 1.43295313 & -0.35599130 \\
H & -4.66803904 & 0.37458745 & -1.66724321 \\
C & -5.02245766 & -0.42014398 & 0.31280777 \\
H & -4.52739271 & -0.40856090 & 1.30103060 \\
H & -5.05758374 & -1.47833747 & -0.00731088 \\
C & -6.44419244 & 0.13362804 & 0.42972437 \\
H & -7.04469059 & -0.44748832 & 1.14874413 \\
H & -6.43065231 & 1.18332936 & 0.77010303 \\
H & -6.95975125 & 0.10369022 & -0.54557862 \\
C & 5.86248375 & 0.57351609 & 0.23840185 \\
H & 6.76174113 & -0.03453088 & 0.06851998 \\
H & 5.79021334 & 0.84608806 & 1.30546199 \\
H & 5.93107784 & 1.51362248 & -0.33562529 \\
\end{tabular}
\end{table}
\begin{table}[h]
\centering
\begin{tabular}{lrrr}
28 &  &  &  \\
\# A4 enol&  &  &  \\
O & 4.67361300 & -1.35693700 & -0.89708100 \\
H & 3.97371400 & -1.29364000 & -1.56405900 \\
C & 4.62987000 & -0.18566900 & -0.17586200 \\
C & 3.33433900 & 0.47798400 & -0.01890100 \\
H & 3.35087200 & 1.48925200 & 0.40608200 \\
C & 2.13599800 & -0.08723000 & -0.35176300 \\
H & 2.11238400 & -1.12033800 & -0.72710000 \\
C & 0.85800800 & 0.58261900 & -0.21129100 \\
H & 0.86480300 & 1.61227700 & 0.17405100 \\
C & -0.33943000 & 0.00426100 & -0.52811200 \\
H & -0.34393900 & -1.02603600 & -0.91319400 \\
C & -1.62378000 & 0.66955700 & -0.38716000 \\
H & -1.61604600 & 1.69908700 & -0.00022500 \\
C & -2.81424600 & 0.09548800 & -0.70711600 \\
H & -2.81062000 & -0.93632700 & -1.09090500 \\
C & -4.15299600 & 0.75478600 & -0.54020200 \\
H & -4.01380300 & 1.81061100 & -0.24221800 \\
H & -4.68827300 & 0.76275300 & -1.51022200 \\
C & -5.03390300 & 0.03843200 & 0.49804600 \\
H & -4.50997600 & 0.04107900 & 1.47135100 \\
H & -5.14649600 & -1.02260000 & 0.20624700 \\
C & -6.41313200 & 0.68588700 & 0.64074000 \\
H & -7.02830800 & 0.16257800 & 1.39112200 \\
H & -6.32203500 & 1.74037900 & 0.95333200 \\
H & -6.95837600 & 0.66634200 & -0.31859300 \\
C & 5.76472400 & 0.28563400 & 0.39849600 \\
H & 5.72459100 & 1.17185800 & 1.03454000 \\
H & 6.71448100 & -0.23440800 & 0.25645600 \\
\end{tabular}
\end{table}

\begin{table}[h]
\centering
\begin{tabular}{lrrr}
26 &  &  &  \\
\# A5 ketone&  &  &  \\
O & 4.62036834 & -1.42038800 & -0.44630421 \\
C & 4.57966951 & -0.21015436 & -0.23868197 \\
C & 3.28830333 & 0.52793884 & -0.11725410 \\
H & 3.31429323 & 1.60933780 & 0.06822191 \\
C & 2.09804403 & -0.12467684 & -0.23377416 \\
H & 2.13803466 & -1.20727299 & -0.41917044 \\
C & 0.80025352 & 0.50600528 & -0.13074318 \\
H & 0.76592879 & 1.58893471 & 0.05471322 \\
C & -0.37613337 & -0.18453451 & -0.25378566 \\
H & -0.33134789 & -1.26766716 & -0.43926318 \\
C & -1.68571028 & 0.42146736 & -0.15503978 \\
H & -1.73197533 & 1.50423784 & 0.03038093 \\
C & -2.85460967 & -0.27874060 & -0.27973414 \\
H & -2.80345972 & -1.36166961 & -0.46505157 \\
C & -4.17458112 & 0.32043380 & -0.18240452 \\
H & -4.22283804 & 1.40331828 & 0.00276237 \\
C & -5.33225430 & -0.38093336 & -0.30724658 \\
H & -5.26602320 & -1.46307812 & -0.49224680 \\
C & -6.70568850 & 0.21731681 & -0.21009704 \\
H & -7.28342897 & -0.24708504 & 0.60895945 \\
H & -6.65648076 & 1.30280093 & -0.02615136 \\
H & -7.27824394 & 0.04894641 & -1.13965974 \\
C & 5.83914741 & 0.62489364 & -0.08984435 \\
H & 6.72697869 & -0.01338355 & -0.19783204 \\
H & 5.84924348 & 1.11753763 & 0.89776484 \\
H & 5.85474059 & 1.42174790 & -0.85333188 \\
\end{tabular}
\end{table}
\begin{table}[h]
\centering
\begin{tabular}{lrrr}
26 &  &  &  \\
\# A5 enol&  &  &  \\
O & 4.57998600 & -1.49318700 & -0.68932800 \\
H & 3.83032400 & -1.56850800 & -1.29861700 \\
C & 4.58342000 & -0.19362200 & -0.23583400 \\
C & 3.29989700 & 0.49955800 & -0.12207300 \\
H & 3.34158600 & 1.57790800 & 0.07546700 \\
C & 2.08270000 & -0.11282800 & -0.23180400 \\
H & 2.03810200 & -1.20191200 & -0.37581200 \\
C & 0.81659900 & 0.58246300 & -0.13745800 \\
H & 0.84451300 & 1.67055200 & 0.01799600 \\
C & -0.39995800 & -0.04024200 & -0.22923400 \\
H & -0.42480500 & -1.12903800 & -0.38414500 \\
C & -1.66893600 & 0.64738300 & -0.13379500 \\
H & -1.64487600 & 1.73576000 & 0.02167500 \\
C & -2.88232100 & 0.02166300 & -0.22559700 \\
H & -2.90445400 & -1.06700700 & -0.38060000 \\
C & -4.15895800 & 0.70878100 & -0.13018700 \\
H & -4.13278400 & 1.79725000 & 0.02479600 \\
C & -5.36289200 & 0.08401200 & -0.22151300 \\
H & -5.37251800 & -1.00484200 & -0.37621600 \\
C & -6.69152200 & 0.77669500 & -0.12534100 \\
H & -7.28876400 & 0.37565100 & 0.71302800 \\
H & -6.56596000 & 1.86082900 & 0.02796300 \\
H & -7.28690500 & 0.62360800 & -1.04318400 \\
C & 5.75682500 & 0.38023100 & 0.13114800 \\
H & 5.76012700 & 1.38288200 & 0.56259700 \\
H & 6.69652200 & -0.16668900 & 0.02900600 \\
\end{tabular}
\end{table}

\begin{table}[h]
\centering
\begin{tabular}{lrrr}
34 &  &  &  \\
\# A7 ketone&  &  &  \\
O & 7.04332496 & -1.47122796 & 0.00474929 \\
C & 8.28837208 & 0.58935444 & 0.00580929 \\
H & 9.16776604 & -0.06941721 & 0.00637870 \\
H & 8.30896615 & 1.24306529 & 0.89481399 \\
H & 8.31062039 & 1.24389999 & -0.88252817 \\
C & 7.01817709 & -0.24277313 & 0.00437199 \\
C & 5.73634498 & 0.52102788 & 0.00254520 \\
H & 5.77572097 & 1.61784164 & 0.00228088 \\
C & 4.53715264 & -0.12738202 & 0.00123543 \\
H & 4.56354681 & -1.22614882 & 0.00158599 \\
C & 3.24892391 & 0.52783215 & -0.00058014 \\
H & 3.22820198 & 1.62692755 & -0.00091547 \\
C & 2.06142791 & -0.15868137 & -0.00186630 \\
H & 2.09219607 & -1.25809518 & -0.00150998 \\
C & 0.76404114 & 0.47193615 & -0.00363434 \\
H & 0.73103201 & 1.57107244 & -0.00401554 \\
C & -0.41928511 & -0.22445223 & -0.00488963 \\
H & -0.38223638 & -1.32381822 & -0.00452890 \\
C & -1.72033683 & 0.39759146 & -0.00670377 \\
H & -1.75887968 & 1.49667598 & -0.00710383 \\
C & -2.90047194 & -0.30319808 & -0.00790420 \\
H & -2.85978645 & -1.40243600 & -0.00747752 \\
C & -4.20584153 & 0.31497767 & -0.00964459 \\
H & -4.24645120 & 1.41397211 & -0.01003273 \\
C & -5.38138512 & -0.38748287 & -0.01080064 \\
H & -5.33920580 & -1.48656161 & -0.01042604 \\
C & -6.69541053 & 0.23073808 & -0.01257856 \\
H & -6.73314829 & 1.32987709 & -0.01310266 \\
C & -7.86100855 & -0.46934525 & -0.01355135 \\
H & -7.80689846 & -1.56785371 & -0.01305662 \\
C & -9.22775883 & 0.15192093 & -0.01514500 \\
H & -9.80766824 & -0.16161154 & 0.87134018 \\
H & -9.16591251 & 1.25228749 & -0.01583096 \\
H & -9.80609605 & -0.16282013 & -0.90221443 \\
\end{tabular}
\end{table}
\begin{table}[h]
\centering
\begin{tabular}{lrrr}
34 &  &  &  \\
\# A7 enol&  &  &  \\
O & -7.13781711 & 1.41840391 & -0.10195208 \\
H & -6.39734926 & 1.68479167 & -0.66746381 \\
C & -8.20925326 & -0.66045231 & 0.21042368 \\
H & -8.15710434 & -1.73821765 & 0.37571349 \\
H & -9.17708500 & -0.15548573 & 0.24218286 \\
C & -7.06974217 & 0.04839716 & 0.00883503 \\
C & -5.75129793 & -0.58280311 & -0.04534960 \\
H & -5.73496427 & -1.67686968 & -0.12508251 \\
C & -4.56785535 & 0.10087697 & 0.01166907 \\
H & -4.58169742 & 1.19197017 & 0.14634950 \\
C & -3.26791810 & -0.52805502 & -0.06141620 \\
H & -3.23696895 & -1.62026833 & -0.18424929 \\
C & -2.08438744 & 0.16132008 & 0.01608267 \\
H & -2.11835672 & 1.25389225 & 0.13974206 \\
C & -0.78445233 & -0.45966842 & -0.05340778 \\
H & -0.74857084 & -1.55197177 & -0.17649657 \\
C & 0.39852054 & 0.23342332 & 0.02547824 \\
H & 0.36094492 & 1.32580015 & 0.14919792 \\
C & 1.69986555 & -0.38387912 & -0.04322256 \\
H & 1.73848021 & -1.47603850 & -0.16711845 \\
C & 2.88067217 & 0.31143352 & 0.03714687 \\
H & 2.84088618 & 1.40365764 & 0.16142598 \\
C & 4.18556330 & -0.30382876 & -0.03127535 \\
H & 4.22503119 & -1.39585350 & -0.15572553 \\
C & 5.36225322 & 0.39221256 & 0.04977489 \\
H & 5.32179801 & 1.48425059 & 0.17452662 \\
C & 6.67533152 & -0.22421039 & -0.01889503 \\
H & 6.71117694 & -1.31638119 & -0.14325038 \\
C & 7.84239497 & 0.46887499 & 0.06138807 \\
H & 7.79060694 & 1.56042624 & 0.18577481 \\
H & 9.78905933 & 0.25965226 & -0.85241756 \\
H & 9.78658716 & 0.05805821 & 0.90970264 \\
C & 9.20785261 & -0.15156063 & -0.00768520 \\
H & 9.14335860 & -1.24464646 & -0.13281651 \\
\end{tabular}
\end{table}

\begin{table}[h]
\centering
\begin{tabular}{lrrr}
42 &  &  &  \\
\# A9 ketone&  &  &  \\
O & 9.47809381 & -1.51456788 & 0.00978680 \\
C & 10.74519626 & 0.53260259 & 0.00424528 \\
H & 11.61749145 & -0.13553444 & 0.00706316 \\
H & 10.77267167 & 1.18887888 & 0.89117263 \\
H & 10.77457633 & 1.18402979 & -0.88617413 \\
C & 9.46619807 & -0.28590749 & 0.00521551 \\
C & 8.19254402 & 0.49145599 & 0.00117481 \\
H & 8.24352966 & 1.58779103 & -0.00254551 \\
C & 6.98637301 & -0.14422490 & 0.00205847 \\
H & 7.00102993 & -1.24320252 & 0.00585440 \\
C & 5.70564417 & 0.52476287 & -0.00167227 \\
H & 5.69675386 & 1.62402551 & -0.00545799 \\
C & 4.51020597 & -0.14886647 & -0.00066079 \\
H & 4.52896604 & -1.24854266 & 0.00314840 \\
C & 3.22092320 & 0.49593967 & -0.00424213 \\
H & 3.19998815 & 1.59539511 & -0.00807846 \\
C & 2.02859561 & -0.18732553 & -0.00318015 \\
H & 2.05337783 & -1.28704365 & 0.00062421 \\
C & 0.73700014 & 0.44916201 & -0.00685060 \\
H & 0.71055139 & 1.54866179 & -0.01066603 \\
C & -0.45377612 & -0.23845125 & -0.00572714 \\
H & -0.42549750 & -1.33810779 & -0.00184761 \\
C & -1.74704951 & 0.39405111 & -0.00929558 \\
H & -1.77625209 & 1.49359630 & -0.01300029 \\
C & -2.93629636 & -0.29570198 & -0.00817702 \\
H & -2.90622350 & -1.39531213 & -0.00445024 \\
C & -4.23169064 & 0.33481614 & -0.01185125 \\
H & -4.26194276 & 1.43429013 & -0.01565566 \\
C & -5.41875191 & -0.35580711 & -0.01064367 \\
H & -5.38781696 & -1.45536655 & -0.00662163 \\
C & -6.71782343 & 0.27402214 & -0.01387315 \\
H & -6.74800766 & 1.37343902 & -0.01711667 \\
C & -7.90084555 & -0.41645379 & -0.01335311 \\
H & -7.87002628 & -1.51591605 & -0.01024834 \\
C & -9.20815181 & 0.21541841 & -0.01702032 \\
H & -9.23401936 & 1.31494447 & -0.02038967 \\
C & -10.38161035 & -0.47164078 & -0.01633754 \\
H & -10.33988375 & -1.57068570 & -0.01296409 \\
C & -11.74130964 & 0.16504368 & -0.01976022 \\
H & -11.66688373 & 1.26465188 & -0.02289234 \\
H & -12.32519119 & -0.13967425 & 0.86722868 \\
H & -12.32308060 & -0.14486959 & -0.90633066 \\
\end{tabular}
\end{table}
\begin{table}[h]
\centering
\begin{tabular}{lrrr}
42 &  &  &  \\
\# A9 enol&  &  &  \\
O & -9.57345422 & 1.45991958 & -0.12446460 \\
H & -8.83222027 & 1.70585036 & -0.69818522 \\
C & -10.66676611 & -0.59975796 & 0.23568474 \\
H & -10.62598945 & -1.67418526 & 0.42428263 \\
H & -11.62890315 & -0.08356260 & 0.25874956 \\
C & -9.52002411 & 0.09198602 & 0.01595929 \\
C & -8.20894044 & -0.55481327 & -0.02740478 \\
H & -8.20483123 & -1.65049442 & -0.08285896 \\
C & -7.01766220 & 0.11691195 & 0.01162614 \\
H & -7.01917526 & 1.21078731 & 0.12219177 \\
C & -5.72546577 & -0.52790768 & -0.05051451 \\
H & -5.70685864 & -1.62283446 & -0.14947536 \\
C & -4.53353785 & 0.14984134 & 0.00941838 \\
H & -4.55511482 & 1.24515795 & 0.10928745 \\
C & -3.24193212 & -0.48697630 & -0.04915883 \\
H & -3.21831654 & -1.58204180 & -0.14865984 \\
C & -2.04978931 & 0.19458579 & 0.01250594 \\
H & -2.07510126 & 1.28973050 & 0.11271511 \\
C & -0.75793991 & -0.43826957 & -0.04524937 \\
H & -0.73115944 & -1.53323216 & -0.14582664 \\
C & 0.43334628 & 0.24595148 & 0.01830652 \\
H & 0.40548289 & 1.34097438 & 0.11940885 \\
C & 1.72636784 & -0.38414258 & -0.03905733 \\
H & 1.75518556 & -1.47900352 & -0.14049052 \\
C & 2.91632928 & 0.30174492 & 0.02551339 \\
H & 2.88710562 & 1.39665868 & 0.12730599 \\
C & 4.21121422 & -0.32707228 & -0.03165716 \\
H & 4.24070352 & -1.42184787 & -0.13367024 \\
C & 5.39906348 & 0.35927399 & 0.03371891 \\
H & 5.36931671 & 1.45413523 & 0.13579607 \\
C & 6.69760393 & -0.26919864 & -0.02311898 \\
H & 6.72681149 & -1.36375013 & -0.12679487 \\
C & 7.88134755 & 0.41694102 & 0.04477852 \\
H & 7.85143472 & 1.51150921 & 0.14854453 \\
C & 9.18810550 & -0.21351999 & -0.01200985 \\
H & 9.21302047 & -1.30820069 & -0.11537404 \\
C & 10.36225246 & 0.46914180 & 0.05497585 \\
H & 10.32161416 & 1.56334598 & 0.15839766 \\
C & 11.72127703 & -0.16646188 & -0.00220753 \\
H & 12.30244204 & 0.05477035 & 0.91092396 \\
H & 11.64549040 & -1.26103965 & -0.10624782 \\
H & 12.30650596 & 0.22232253 & -0.85476325 \\
\end{tabular}
\end{table} 

\begin{table}[h]
\centering
\begin{tabular}{lrrr}
58 &  &  &  \\
\# A13 ketone&  &  &  \\
O   &     14.39419293   &     -1.55257919  &  -0.06344859  \\
C   &     15.67961815   &     0.48224111   &  0.00255222   \\
H   &     15.71357468   &     1.16378833   &  -0.86480286  \\
H   &     15.71372299   &     1.10718325   &  0.91150406   \\
H   &     16.54620363   &     -0.19295434  &  -0.01898366  \\
C   &     14.39353682   &     -0.32452304  &  -0.02331818  \\
C   &     13.12667990   &     0.46388621   &  0.00252788   \\
H   &     13.18745786   &     1.55916484   &  0.03744192   \\
C   &     11.91497632   &     -0.16092769  &  -0.01678095  \\
H   &     11.91993459   &     -1.25942957  &  -0.05190776  \\
C   &     10.64033170   &     0.51903635   &  0.00576645   \\
H   &     10.64127448   &     1.61779273   &  0.04075604   \\
C   &     9.43869338    &     -0.14350569  &  -0.01450820  \\
H   &     9.44741644    &     -1.24274804  &  -0.04953331  \\
C   &     8.15581152    &     0.51281177   &  0.00728377   \\
H   &     8.14510397    &     1.61187681   &  0.04228312   \\
C   &     6.95662485    &     -0.15895427  &  -0.01325484  \\
H   &     6.97098098    &     -1.25830530  &  -0.04827268  \\
C   &     5.67208856    &     0.48953519   &  0.00830853   \\
H   &     5.65620601    &     1.58871773   &  0.04332125   \\
C   &     4.47363305    &     -0.18621009  &  -0.01235371  \\
H   &     4.49114160    &     -1.28554629  &  -0.04737618  \\
C   &     3.18849120    &     0.45862343   &  0.00910431   \\
H   &     3.17012643    &     1.55784991   &  0.04412463   \\
C   &     1.99037945    &     -0.21903588  &  -0.01161708  \\
H   &     2.00951857    &     -1.31834949  &  -0.04664436  \\
C   &     0.70488679    &     0.42392580   &  0.00978718   \\
H   &     0.68527333    &     1.52316764   &  0.04481343   \\
C   &     -0.49295961   &     -0.25469765  &  -0.01096380  \\
H   &     -0.47296539   &     -1.35399956  &  -0.04599455  \\
C   &     -1.77882112   &     0.38730832   &  0.01041781   \\
H   &     -1.79902143   &     1.48655697   &  0.04544947   \\
C   &     -2.97636253   &     -0.29172250  &  -0.01034958  \\
H   &     -2.95600148   &     -1.39102037  &  -0.04539159  \\
C   &     -4.26276146   &     0.34989424   &  0.01102684   \\
H   &     -4.28313736   &     1.44914533   &  0.04604672   \\
C   &     -5.45980000   &     -0.32917959  &  -0.00972462  \\
H   &     -5.43943293   &     -1.42848172  &  -0.04473314  \\
C   &     -6.74714745   &     0.31239635   &  0.01160321   \\
H   &     -6.76733092   &     1.41164020   &  0.04660876   \\
C   &     -7.94325767   &     -0.36645813  &  -0.00925194  \\
H   &     -7.92316172   &     -1.46575022  &  -0.04420145  \\
C   &     -9.23219845   &     0.27562930   &  0.01197916   \\
H   &     -9.25175071   &     1.37484854   &  0.04714863   \\
C   &     -10.42671250  &     -0.40244624  &  -0.00913316  \\
H   &     -10.40719645  &     -1.50171782  &  -0.04429499  \\
C   &     -11.71893485  &     0.24066793   &  0.01205424   \\
H   &     -11.73706581  &     1.33982917   &  0.04725339   \\
C   &     -12.90980448  &     -0.43620574  &  -0.00908854  \\
H   &     -12.89140558  &     -1.53538415  &  -0.04428973  \\
C   &     -14.20981989  &     0.21014961   &  0.01218402   \\
H   &     -14.22284480  &     1.30937492   &  0.04737681   \\
C   &     -15.39143538  &     -0.46257491  &  -0.00882239  \\
H   &     -15.36289715  &     -1.56148113  &  -0.04401389  \\
C   &     -16.74347439  &     0.19005895   &  0.01267414   \\
H   &     -17.33043182  &     -0.08153420  &  -0.88301591  \\
H   &     -16.65568101  &     1.28813542   &  0.04778372   \\
H   &     -17.32963663  &     -0.13827739  &  0.88968710   \\
\end{tabular}
\end{table}
\begin{table}[h]
\centering
\begin{tabular}{lrrr}
58 &  &  &  \\
\# 13 enol&  &  &  \\
O &   -14.47603994 &  1.52565223 & -0.25389996 \\
H &   -13.73492619 &  1.71634422 & -0.84846102 \\
C &   -15.59379750 & -0.48576055 &  0.26790491 \\    
H &   -15.56579756 & -1.54230799 &  0.54101134 \\
H &   -16.54942177 &  0.04262266 &  0.25134814 \\ 
C &   -14.43906088 &  0.17225794 & -0.00690558 \\
C &   -13.13614272 & -0.49202041 & -0.00019508 \\
H &   -13.14524306 & -1.58860168 &  0.03132182 \\ 
C &   -11.93673754 &  0.16638525 & -0.01554756 \\  
H &   -11.92537313 &  1.26548606 &  0.00812293 \\
C &   -10.65265854 & -0.49681141 & -0.02754152 \\
H &   -10.64722139 & -1.59628551 & -0.03938679 \\
C &    -9.45239710 &  0.16926381 & -0.02252892 \\
H &    -9.46075062 &  1.26922033 & -0.00963146 \\
C &    -8.16906565 & -0.48573415 & -0.03150757 \\
H &    -8.15864405 & -1.58545346 & -0.04427337 \\
C &    -6.96826596 &  0.18427326 & -0.02449020 \\
H &    -6.98036487 &  1.28413163 & -0.01089103 \\
C &    -5.68511675 & -0.46661734 & -0.03282284 \\
H &    -5.67143638 & -1.56635409 & -0.04701169 \\
C &    -4.48464526 &  0.20634273 & -0.02355785 \\
H &    -4.49949211 &  1.30616245 & -0.00878272 \\
C &    -3.20119862 & -0.44168146 & -0.03147153 \\
H &    -3.18538642 & -1.54141485 & -0.04684114 \\
C &    -2.00114360 &  0.23319718 & -0.02053177 \\
H &    -2.01771627 &  1.33299273 & -0.00475194 \\
C &    -0.71728183 & -0.41301384 & -0.02805158 \\
H &    -0.70011547 & -1.51273912 & -0.04439102 \\
C &     0.48233278 &  0.26306774 & -0.01568488 \\
H &     0.46466231 &  1.36284212 &  0.00098202 \\
C &     1.76672344 & -0.38196402 & -0.02288328 \\
H &     1.78479346 & -1.48167318 & -0.04005935 \\
C &     2.96577084 &  0.29492367 & -0.00931170 \\
H &     2.94735642 &  1.39467477 &  0.00812735 \\
C &     4.25093455 & -0.34927716 & -0.01629039 \\
H &     4.26966205 & -1.44896376 & -0.03414623 \\
C &     5.44912049 &  0.32821151 & -0.00179491 \\
H &     5.43017299 &  1.42794978 &  0.01627705 \\
C &     6.73555775 & -0.31533532 & -0.00863915 \\
H &     6.75458427 & -1.41498715 & -0.02707110 \\
C &     7.93249026 &  0.36248453 &  0.00663004 \\
H &     7.91346077 &  1.46217275 &  0.02521010 \\
C &     9.22068953 & -0.28124839 & -0.00010525 \\
H &     9.23937909 & -1.38086181 & -0.01921012 \\
C &    10.41583235 &  0.39602534 &  0.01597730 \\
H &    10.39708526 &  1.49569117 &  0.03517523 \\
C &    11.70745546 & -0.24842678 &  0.00931244 \\
H &    11.72499097 & -1.34796226 & -0.01012299 \\
C &    12.89872105 &  0.42799413 &  0.02574788 \\
H &    12.88076129 &  1.52754832 &  0.04525587 \\
C &    14.19835376 & -0.21931364 &  0.01905179 \\
H &    14.21108146 & -1.31890052 & -0.00062941 \\
C &    15.38015724 &  0.45326801 &  0.03559413 \\
H &    15.35188134 &  1.55254969 &  0.05524806 \\
C &    16.73189856 & -0.20021177 &  0.02896641 \\
H &    17.32198414 &  0.11555338 & -0.85001893 \\
H &    16.64360325 & -1.29861013 &  0.00874872 \\
H &    17.31517350 &  0.08345804 &  0.92330794 \\
\end{tabular}
\end{table} 

\clearpage 
\subsection{Difficult set}

\begin{table}[h]
\centering
\begin{tabular}{lrrr}
23 &  &  &  \\
\# B5 ketone&  &  &  \\
O & -5.18796110 & -1.47625967 & -0.06081888 \\
C & -5.09183417 & -0.26900208 & -0.24978337 \\
H & -6.00522694 & 0.35805732 & -0.40470191 \\
C & -3.80717251 & 0.46891694 & -0.29302913 \\
H & -3.83691287 & 1.55133526 & -0.46930269 \\
C & -2.62084518 & -0.18083366 & -0.11849185 \\
H & -2.66371801 & -1.26553646 & 0.05384649 \\
C & -1.32242831 & 0.45451386 & -0.14447053 \\
H & -1.28138623 & 1.53922257 & -0.31678709 \\
C & -0.15282983 & -0.23566166 & 0.03561418 \\
H & -0.20561099 & -1.32074855 & 0.20732549 \\
C & 1.15885538 & 0.37281091 & 0.01477635 \\
H & 1.21416536 & 1.45737249 & -0.15663514 \\
C & 2.31967806 & -0.32857875 & 0.19611293 \\
H & 2.25894044 & -1.41335278 & 0.36726628 \\
C & 3.64252976 & 0.27139763 & 0.17730607 \\
H & 3.70081900 & 1.35602968 & 0.00614807 \\
C & 4.79139238 & -0.43230743 & 0.35822217 \\
H & 4.71470745 & -1.51623146 & 0.52819159 \\
C & 6.16824936 & 0.16545639 & 0.34288024 \\
H & 6.79027695 & -0.29265251 & -0.44671426 \\
H & 6.13001101 & 1.25251929 & 0.16601763 \\
H & 6.68788474 & -0.01186895 & 1.30139809 \\
\end{tabular}
\end{table}
\begin{table}[h]
\centering
\begin{tabular}{lrrr}
23 &  &  &  \\
\# B5 enol&  &  &  \\
O & -6.30027500 & 0.26663200 & -0.40724700 \\
H & -6.20969000 & 1.22107700 & -0.55570700 \\
C & -5.04979400 & -0.23992500 & -0.25199300 \\
H & -5.07615300 & -1.32130100 & -0.07931700 \\
C & -3.87852900 & 0.45055800 & -0.29427300 \\
H & -3.89791400 & 1.53562000 & -0.47024100 \\
C & -2.59152900 & -0.19155300 & -0.11503900 \\
H & -2.59165800 & -1.27758000 & 0.05993700 \\
C & -1.38822400 & 0.45977800 & -0.14911600 \\
H & -1.37597400 & 1.54545300 & -0.32336100 \\
C & -0.11499200 & -0.20087400 & 0.03236200 \\
H & -0.13078900 & -1.28714900 & 0.20648600 \\
C & 1.09669000 & 0.43851600 & 0.00071100 \\
H & 1.11768400 & 1.52427400 & -0.17298700 \\
C & 2.36491300 & -0.23298600 & 0.18356900 \\
H & 2.33925500 & -1.31903600 & 0.35692100 \\
C & 3.57889200 & 0.39660500 & 0.15373900 \\
H & 3.61065100 & 1.48213300 & -0.01920900 \\
C & 4.84828200 & -0.29000800 & 0.33914000 \\
H & 4.80438500 & -1.37429500 & 0.51120000 \\
C & 6.05588300 & 0.32818700 & 0.31073600 \\
H & 6.98474600 & -0.23069000 & 0.45555000 \\
H & 6.13225200 & 1.40808100 & 0.14125600 \\
\end{tabular}
\end{table}

\begin{table}[h]
\centering
\begin{tabular}{lrrr}
31 &  &  &  \\
\# B7 ketone&  &  &  \\
O & 7.72315523 & -1.30164624 & -0.00002700 \\
C & 7.63351251 & -0.07908928 & -0.00001891 \\
H & 8.55186396 & 0.55974680 & -0.00001328 \\
C & 6.35083845 & 0.66296155 & -0.00001841 \\
H & 6.38629594 & 1.75949308 & -0.00001640 \\
C & 5.15825602 & -0.00003495 & -0.00001849 \\
H & 5.19521125 & -1.09856932 & -0.00001919 \\
C & 3.86268483 & 0.63841648 & -0.00001624 \\
H & 3.82752542 & 1.73699785 & -0.00001705 \\
C & 2.68489367 & -0.06508433 & -0.00001165 \\
H & 2.73135510 & -1.16399766 & -0.00000994 \\
C & 1.37881261 & 0.54637931 & -0.00000753 \\
H & 1.32910802 & 1.64480393 & -0.00001059 \\
C & 0.20643638 & -0.16854206 & 0.00000097 \\
H & 0.26084370 & -1.26720592 & 0.00000469 \\
C & -1.10411481 & 0.43270051 & 0.00000640 \\
H & -1.16055068 & 1.53095637 & 0.00000133 \\
C & -2.27249569 & -0.28766594 & 0.00001826 \\
H & -2.21340901 & -1.38607661 & 0.00002392 \\
C & -3.58799086 & 0.30843015 & 0.00002452 \\
H & -3.64740216 & 1.40654068 & 0.00001790 \\
C & -4.75120646 & -0.41436335 & 0.00003893 \\
H & -4.68988326 & -1.51254327 & 0.00004605 \\
C & -6.07574561 & 0.18082336 & 0.00004714 \\
H & -6.13287784 & 1.27909628 & 0.00003587 \\
C & -7.22869661 & -0.53994147 & 0.00006563 \\
H & -7.15495783 & -1.63730897 & 0.00007355 \\
C & -8.60628352 & 0.05677016 & 0.00008753 \\
H & -9.17954279 & -0.26776959 & 0.88691379 \\
H & -8.56424215 & 1.15805363 & -0.00001029 \\
H & -9.17964279 & -0.26792741 & -0.88661487 \\
\end{tabular}
\end{table}
\begin{table}[h]
\centering
\begin{tabular}{lrrr}
31 &  &  &  \\
\# B7 enol&  &  &  \\
O & -7.75540013 & 1.16640290 & -0.00697816 \\
H & -6.91614493 & 1.65092749 & 0.00168443 \\
C & -7.46872084 & -0.16164016 & -0.02698635 \\
H & -8.38387989 & -0.76085402 & -0.03699010 \\
C & -6.23722380 & -0.74660183 & -0.03441440 \\
H & -6.21708495 & -1.84117055 & -0.05073622 \\
C & -4.97219786 & -0.03994099 & -0.02246877 \\
H & -4.97498013 & 1.06129519 & -0.00654782 \\
C & -3.74520445 & -0.65089761 & -0.02964377 \\
H & -3.70392579 & -1.74960617 & -0.04551286 \\
C & -2.49136844 & 0.06340107 & -0.01752824 \\
H & -2.53359038 & 1.16285004 & -0.00183696 \\
C & -1.25998469 & -0.54391833 & -0.02437224 \\
H & -1.21470921 & -1.64274596 & -0.04005619 \\
C & -0.00997416 & 0.17391995 & -0.01219938 \\
H & -0.05612927 & 1.27295420 & 0.00342330 \\
C & 1.22339676 & -0.43067611 & -0.01892459 \\
H & 1.27143269 & -1.52943272 & -0.03454129 \\
C & 2.47177819 & 0.29076600 & -0.00668862 \\
H & 2.42203766 & 1.38957231 & 0.00891004 \\
C & 3.70557630 & -0.31031394 & -0.01334366 \\
H & 3.75733856 & -1.40881518 & -0.02894755 \\
C & 4.95418111 & 0.41673365 & -0.00101314 \\
H & 4.89869369 & 1.51518771 & 0.01456770 \\
C & 6.18684878 & -0.17784508 & -0.00756126 \\
H & 6.24685938 & -1.27579115 & -0.02314501 \\
C & 7.43921629 & 0.56171986 & 0.00497929 \\
H & 7.36775743 & 1.65792902 & 0.02045167 \\
H & 9.57984231 & 0.57361688 & 0.00867816 \\
C & 8.66367973 & -0.02356945 & -0.00135901 \\
H & 8.76688937 & -1.11436423 & -0.01679764 \\
\end{tabular}
\end{table}

\begin{table}[h]
\centering
\begin{tabular}{lrrr}
39 &  &  &  \\
\# B9 ketone&  &  &  \\
O & 10.16296679 & -1.39272295 & -0.00017086 \\
C & 10.09551449 & -0.16872422 & -0.00005341 \\
H & 11.02524348 & 0.45340509 & 0.00000982 \\
C & 8.82644995 & 0.59632186 & 0.00001214 \\
H & 8.88166927 & 1.69204312 & 0.00011311 \\
C & 7.62192433 & -0.04506915 & -0.00005045 \\
H & 7.63896043 & -1.14409312 & -0.00015114 \\
C & 6.33856414 & 0.61680685 & 0.00000674 \\
H & 6.32342304 & 1.71586319 & 0.00010533 \\
C & 5.14752163 & -0.06504912 & -0.00005552 \\
H & 5.17376365 & -1.16463181 & -0.00015415 \\
C & 3.85405312 & 0.57040516 & 0.00000129 \\
H & 3.82465489 & 1.66959295 & 0.00009709 \\
C & 2.66722759 & -0.12263538 & -0.00005691 \\
H & 2.70109889 & -1.22213859 & -0.00015284 \\
C & 1.37058625 & 0.50299653 & 0.00000083 \\
H & 1.33456572 & 1.60217896 & 0.00009447 \\
C & 0.18601399 & -0.19539849 & -0.00005352 \\
H & 0.22432748 & -1.29477377 & -0.00014707 \\
H & -1.15236909 & 1.52430978 & 0.00009758 \\
C & -1.11288417 & 0.42516535 & 0.00000541 \\
C & -2.29546554 & -0.27605082 & -0.00004654 \\
H & -2.25470253 & -1.37532809 & -0.00013870 \\
C & -3.59688538 & 0.34179866 & 0.00001266 \\
H & -3.63805917 & 1.44089306 & 0.00010325 \\
C & -4.77695147 & -0.36079188 & -0.00003667 \\
H & -4.73488181 & -1.45998828 & -0.00012658 \\
H & -6.12374489 & 1.35489284 & 0.00011676 \\
C & -6.08236262 & 0.25586245 & 0.00002186 \\
C & -7.25816212 & -0.44690830 & -0.00003512 \\
H & -7.21587060 & -1.54598769 & -0.00012983 \\
C & -8.57197481 & 0.17129637 & 0.00002237 \\
H & -8.60940016 & 1.27047756 & 0.00011352 \\
C & -9.73809418 & -0.52816850 & -0.00002923 \\
H & -9.68469100 & -1.62671312 & -0.00012005 \\
C & -11.10446745 & 0.09402734 & 0.00002676 \\
H & -11.04177045 & 1.19435770 & 0.00013304 \\
H & -11.68392165 & -0.21964896 & 0.88677685 \\
H & -11.68391549 & -0.21947727 & -0.88678833 \\
\end{tabular}
\end{table}
\begin{table}[h]
\centering
\begin{tabular}{lrrr}
39 &  &  &  \\
\# B9 enol&  &  &  \\
O & -10.19762000 & 1.24603000 & -0.01329000 \\
H & -9.35063000 & 1.71691000 & -0.00472000 \\
C & -9.93255000 & -0.08640000 & -0.03227000 \\
H & -10.85739000 & -0.67076000 & -0.04245000 \\
C & -8.71059000 & -0.69134000 & -0.03944000 \\
H & -8.70832000 & -1.78609000 & -0.05602000 \\
C & -7.43445000 & -0.00546000 & -0.02710000 \\
H & -7.41913000 & 1.09567000 & -0.01122000 \\
C & -6.21727000 & -0.63649000 & -0.03406000 \\
H & -6.19414000 & -1.73574000 & -0.04983000 \\
C & -4.95242000 & 0.05689000 & -0.02195000 \\
H & -4.97630000 & 1.15689000 & -0.00638000 \\
C & -3.73054000 & -0.57089000 & -0.02861000 \\
H & -3.70381000 & -1.67035000 & -0.04425000 \\
C & -2.46990000 & 0.12550000 & -0.01628000 \\
H & -2.49715000 & 1.22518000 & -0.00070000 \\
C & -1.24547000 & -0.50021000 & -0.02284000 \\
H & -1.21679000 & -1.59970000 & -0.03843000 \\
C & 0.01285000 & 0.19848000 & -0.01050000 \\
H & -0.01655000 & 1.29807000 & 0.00507000 \\
C & 1.23860000 & -0.42532000 & -0.01701000 \\
H & 1.26905000 & -1.52477000 & -0.03257000 \\
C & 2.49591000 & 0.27552000 & -0.00465000 \\
H & 2.46461000 & 1.37502000 & 0.01089000 \\
C & 3.72198000 & -0.34637000 & -0.01111000 \\
H & 3.75439000 & -1.44573000 & -0.02666000 \\
C & 4.97937000 & 0.35716000 & 0.00132000 \\
H & 4.94545000 & 1.45654000 & 0.01686000 \\
C & 6.20485000 & -0.26196000 & -0.00509000 \\
H & 6.24039000 & -1.36111000 & -0.02061000 \\
C & 7.46354000 & 0.44654000 & 0.00736000 \\
H & 7.42443000 & 1.54567000 & 0.02286000 \\
C & 8.68742000 & -0.16656000 & 0.00103000 \\
H & 8.73085000 & -1.26529000 & -0.01451000 \\
C & 9.95063000 & 0.55399000 & 0.01375000 \\
H & 9.89593000 & 1.65114000 & 0.02922000 \\
H & 12.09119000 & 0.53322000 & 0.01763000 \\
C & 11.16586000 & -0.04992000 & 0.00757000 \\
H & 11.25230000 & -1.14217000 & -0.00786000 \\
\end{tabular}
\end{table}

\end{document}